\documentclass[aps,pre,twocolumn,superscriptaddress,superscriptreference]{revtex4-2}

\usepackage{amsmath,bbold,bm,amssymb,scalerel,mathtools}
\usepackage{graphicx}
\usepackage{color}
\usepackage{enumitem}
\usepackage{algorithm,algpseudocode}
\usepackage{multirow}
\usepackage{colortbl,booktabs}
\usepackage{placeins}
\usepackage[usenames,dvipsnames]{xcolor}

\usepackage[colorlinks, linkcolor=BrickRed, urlcolor=blue!50!black, citecolor=blue!50!black]{hyperref}

\newcommand\equalhat{%
	\let\savearraystretch\arraystretch
	\renewcommand\arraystretch{0.3}
	\begin{array}{c}
		\stretchto{
			\scalerel*[\widthof{=}]{\wedge}
			{\rule{1ex}{3ex}}%
		}{0.5ex}\\ 
		=%
	\end{array}
	\let\arraystretch\savearraystretch
}

\newcommand\diff{\mathrm{d}}

\newcommand{\gj}[1]{\textcolor{black}{#1}}

\usepackage[normalem]{ulem}

\begin{document}

\title{Structural properties of liquids in extreme confinement}

\author{Gerhard Jung}
\email{jung.gerhard@umontpellier.fr}
\affiliation{Institut f\"ur Theoretische Physik, Universit\"at Innsbruck, Technikerstra{\ss}e 21A, 6020 Innsbruck, Austria.}
\affiliation{Laboratoire Charles Coulomb (L2C), Université de Montpellier, CNRS, 34095 Montpellier, France.}

\author{Thomas Franosch}
\email{thomas.franosch@uibk.ac.at}
\affiliation{Institut f\"ur Theoretische Physik, Universit\"at Innsbruck, 
Technikerstra{\ss}e 21A, 6020 Innsbruck, Austria.}

\date{\today}

\begin{abstract}

We simulate a hard-sphere liquid in confined geometry where the separation of the two parallel, hard walls is smaller than two particle diameters. By systematically reducing the wall separation we analyze the behavior of structural and thermodynamic properties, such as inhomogeneous density profiles, structure factors, and compressibilities when approaching the two-dimensional limit. In agreement with asymptotic predictions, we find for quasi-two-dimensional fluids that the density profile becomes parabolic and the structure factor converges towards its two-dimensional counterpart. To extract the compressibility in polydisperse samples a perturbative expression is used which qualitatively influences the observed non-monotonic dependence of the compressibility with wall separation. We also present theoretical calculations based on fundamental{-}measure theory and integral{-}equation theory, which are in very good agreement with the simulation results. 

\end{abstract}

\maketitle

\section{Introduction}

Simple liquids in confinement exhibit a rich phenomenology, including inhomogeneous density profiles and layering \cite{Evans_1990,Noyola1991,Nemeth:PRE59:1999,Mittal:PRL100:2008}, anisotropic structure factors and orientational alignment \cite{Nygard2009,Nygard2016,Mandal2017a}, as well as multiple-reentrant crystallization and glass transitions \cite{exp:Pieranski1983,theo:Schmidt1996,theo:Schmidt1997,Alba_Simionesco_2006,Mandal2014,varnik2016non}, and demixing \cite{Jung2020_cryst}. All these phenomena become particularly pronounced 
in the regime of strong confinement where the wall separation extends to less than three particle diameters. The structure of simple liquids in 
this regime has been extensively studied in theory \cite{antonchenko1984nature,theo:Schmidt1996,theo:Schmidt1997,doi:10.1098/rspa.2007.0115,PhysRevE.78.011602,PhysRevLett.109.240601,Lang2014a,Mandal2014}, simulations \cite{doi:10.1063/1.457334,theo:Schmidt1996,theo:Schmidt1997,henderson1997second,alejandre1996effect,Fortini_2006,doi:10.1063/1.481430,doi:10.1063/1.3623783,Mandal2014,C4SM00125G,Mandal2017,Mandal2017a,Jung2020_cryst,roberts2020dynamics}, and experiments \cite{Nygard2009,Nygard2012,O_uz_2012,Nygard2013,Nygard2016,weiss2019structure,Liu2020}. These studies highlight that basically all of the above phenomena can be explained by the interplay between the length scale of the confinement and the typical particle diameters and interaction range. The non-monotonic dependence of the inhomogeneous density profile and of the generalized structure factor \gj{on the wall separation} as reported by theory and simulations in Refs.~\cite{Mandal2014,Mandal2017a} can for example be explained with the concepts of commensurate packing if the confinement length is close to an integer multiple of the particle diameter, and incommensurate packing otherwise. The former favors the packing into pronounced layers 
with few particles between, while the latter inevitably requires some particles to be placed between those layers, creating a somewhat ``frustrated'' and interlocked liquid structure \cite{Mandal2014,Nygard2016}. When increasing the packing fraction the same effect leads to the emergence of reentrant crystallization transitions where an originally crystalline sample melts upon reduction of the wall separation when entering the incommensurate regime, and freezes again when reducing the wall separation even further \cite{exp:Pieranski1983,theo:Schmidt1996,theo:Schmidt1997,Alba_Simionesco_2006}. The reason is that incommensurate packing disfavors freezing into stacked layers of two-dimensional crystals and thus significantly increases the critical packing fraction.

Observing these phenomena in strong confinement immediately suggests the question how the structure of a three-dimensional liquid behaves upon even more extreme confinement and how it finally converges towards the structure of its 2D counterpart. Such questions are particularly interesting to understand the dimensional crossover behavior, for example for the emergence of the hexatic phase \cite{doi:10.1063/1.3623783,C4SM00125G}, or the decoupling of 
lateral and transverse currents \cite{Schilling2016,Mandal2017}. Important results have already been obtained from theory and simulations for the inhomogeneous density profiles of hard-sphere liquids, highlighting the emergence of parabolic profiles in the regime of extreme confinement, which become increasingly flat and finally converge towards the area density of the hard-disk fluid \cite{antonchenko1984nature,Lang2014a}.  Similarly, the convergence of the critical packing fraction for the freezing transition in extreme confinement towards its 2D counterpart has been quantified in computer simulations \cite{theo:Schmidt1996,theo:Schmidt1997} in very good agreement with density-functional theory and asymptotic formulas \cite{theo:Schmidt1997,doi:10.1098/rspa.2007.0115,PhysRevLett.109.240601}. 

In contrast to these well-established results, very little is known about the actual packing of the particles as characterized for example by the anisotropic structure factor \cite{Mandal2017a,Nygard2016}.  The particular focus of this manuscript therefore lies on studying the impact of confinement on packing and thermodynamic 
quantities of hard-sphere fluids and how these properties \gj{converge} towards the quasi-two-dimensional limit in extreme confinement. For \gj{monodisperse} systems, theoretical predictions have been derived for the rapidity of convergence of the density profiles and structure factors towards their 2D counterparts \cite{Lang2014a}, thus we also compare simulation and numerical results performed in this work with these theoretically predicted asymptotic formulas. \gj{Since experimental studies are usually based on polydisperse samples \cite{Nygard2009,Nygard2012,Nygard2013,Nygard2016} we also analyze the behavior of polydisperse hard spheres in strong confinement, where we find an anomalous behavior of the structure factor at small wavelengths which could be connected to the observation of microscopic demixing of differently sized particles. To extract the compressibility of the polydisperse samples from the structure factor we adapt the perturbative expansion proposed in Ref.~\cite{Berthier2011}.} 

 Our manuscript is organized as follows. In Section \ref{sec:model} we introduce the confined geometry and the simulation model. We then discuss in Section \ref{sec:structure_intro} the structural quantities used to characterize the confined fluid and present numerical algorithms to calculate these quantities theoretically. Afterwards, we recapitulate the results presented in Ref.~\cite{Lang2014a} on the convergence of these structural quantities towards the 2D fluid in Section \ref{sec:theory}. The simulation 
results compared to the theory for the density profiles and structure factors are then presented and discussed in Sections \ref{sec:density} and \ref{sec:structure}, respectively. We summarize and conclude in Section \ref{sec:conclusions}.
  
  \section{Slit geometry and simulation model}
  \label{sec:model}

  \begin{figure}[b]
  	\includegraphics[scale=0.27]{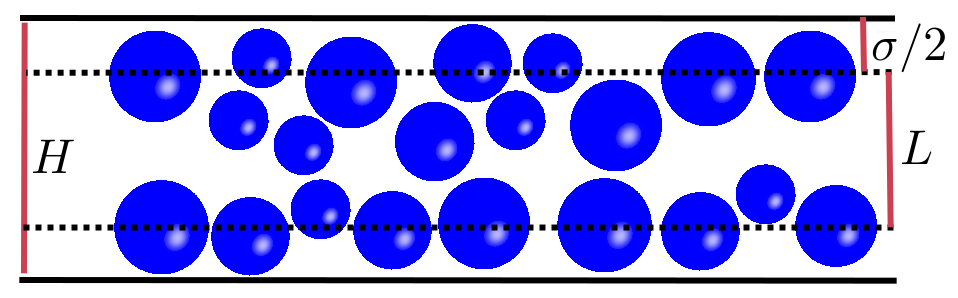}
  	\caption{Schematic of the slit geometry highlighting the important length scales: the wall separation $ H $, the accessible slit width $ L $ and the average particle diameter $ {\sigma}. $  }
  	\label{fig:slit_schematic}
  \end{figure}
  
\gj{The systems considered} in this work are comprised of $ N $ polydisperse \gj{or monodisperse} hard spheres with \emph{average} particle diameter $ {\sigma} $, confined between two parallel, hard walls at a distance $ H $ (see Fig.~\ref{fig:slit_schematic}). In the following, we will denote the direction perpendicular to these walls as the transverse direction ($ z $-direction) and the directions parallel to the walls as the lateral directions. Due to the finite particle size, the centers of particles with diameter $ {\sigma} $ cannot enter the wall regions which leads to the definition of an accessible slit width $ L = H - {\sigma} $, which will play an important role since 
it is the quantity that approaches zero in the quasi-two-dimensional limit.

We will consider two control parameters in this work. First, the packing fraction of the particles in the slit, which is defined as $ \varphi = V_\text{sphere}/V$ where $ V_\text{sphere} $ is the total volume occupied 
by the spheres and $ V = A H $ is the total volume of the slit with area $ A. $ The packing fraction is directly related to the number density $ n_\text{3D} = N/V= 6 \varphi/\pi \sigma^3 $. Second, we consider 
the area number density $ n_0= N/A $ which will prove to be useful for the convergence analysis towards to the 2D limit \cite{Lang2014a}. It should be noted that since $\varphi = \pi \sigma^3 n_0  / 6(L+\sigma) $ we have $\varphi \to \pi \sigma^2 n_0 / 6 $ for $L\to 0$, thus also $\varphi$ becomes constant for small wall separation.

We will study systems with different polydispersities $ \delta $, defined 
as the standard deviation divided by the mean $ \sigma $ of the Gaussian distribution from which the particle diameters are drawn at initialization. In particular, we will also focus on the monodisperse case, which allows analyzing systematically the limit $ L\rightarrow 0. $ After initialization of $ N = 5000 - 13000 $ particles, dense  packings  are created using a linear compression algorithm \cite{Woodcock1981_compress1,stillinger1993crystalline}. The system is then equilibrated using event-driven molecular dynamics (EDMD) simulations \cite{edmd:alder1957,edmd:RAPAPORT1980,edmd:bannerman2011} for at least $ 10^{10} $ events, which is sufficient to ensure full equilibration. Both the compression algorithm as well as the EDMD simulations are implemented in the open-source library DynamO \cite{edmd:bannerman2011} which has thus been used for all simulations presented in this work.

\section{Structural quantities and Theoretical models }
\label{sec:structure_intro}

The structural quantities of interest for this work are the inhomogeneous 
density profile $ n(z) $, the generalized structure factor $ S_{\mu \nu}(q) $ and the anisotropic structure factor $ S(q,q_\perp) $. The inhomogeneous density profile $ n(z) $ is defined as the  average of the particle density, 
\begin{align}\label{key}
n(z)&= \left\langle \rho(\bm{r},z,t) \right\rangle,\\
\rho(\bm{r},z,t)& := \sum_{n=1}^{N} \delta \left[\bm{r} - \bm{r}_n(t)\right] \delta\left[ z-z_n(t) \right] , 
\end{align}
with the particle positions at time $t$ in lateral $ \bm{r}_n(t)=(x_n(t),y_n(t)) $ and transverse direction $ z_n(t) $. Here and in the following $\langle \ldots \rangle$ indicates a \gj{time average in the computer simulations}.
The profile only depends on the transverse direction $ z $ due to translational invariance in the lateral dimensions.

In addition to the computer simulations we will  report results obtained from fundamental{-}measure theory (FMT) for polydisperse particles in 
confinement \cite{fmt:Roth2010,Mandal2014,Jung2020_cryst}. Fundamental{-}measure theory is based 
on density{-}functional theory and the underlying idea is to minimize 
the grand potential. This minimization procedure leads to the fundamental 
equation \cite{fmt:Rosenfeld1989,fmt:Roth2010},
\begin{equation}\label{eq:fmt}
\ln \left( n_i(z) \lambda_i^3 \right) = \beta \mu_i - \beta \frac{\delta \mathcal{F}^\text{ex}[n_i]}{\delta n_i(z)} - \beta V_i(z),
\end{equation} 
for the inhomogeneous number densities $ n_i(z) $ of different species $ i $. Here, $ \lambda_i $ denotes the thermal wavelength, \gj{which does not affect the structural properties.} $ \mu_i $ is the chemical potential, $ \beta = (k_BT)^{-1} $ is the inverse temperature and $ V_i(z) $ is an external potential which will model the hard walls. The excess free energy $ \mathcal{F}^\text{ex}[n_i] $ contains the contribution of the particle interactions and has to be approximated in FMT. In this work we adapt the White-Bear Version II functional \cite{fmt:Roth2010}. The 
different species emulate the polydispersity of the system, where the diameter of each species is given by $ \sigma_i = \sigma_\text{min} + (i+0.5) \Delta \sigma$ for  $ i=0,\hdots,m-1   $. We use $ m = 25 $ with $ \Delta \sigma = 0.024\,\sigma $ and $ \sigma_\text{min} = 0.7\,\sigma $, similar to Ref.~\cite{Jung2020_cryst}. The density profile is determined by a self-consistent iterative procedure, as described in Refs.~\cite{fmt:Roth2010,Jung2020_cryst}. The definition of the inhomogeneous number density $ n_i(z) $ in 
FMT is consistent with the density profile $ n(z) $ defined above, which can be evaluated as $ n(z) = \sum_i n_i(z) $. In the monodisperse case, 
the algorithm significantly simplifies, featuring only a single species with diameter $ \sigma_i = \sigma. $  

The generalized structure factor (GSF) can be calculated from the fluctuating density modes \cite{Lang2010,Lang2012,Mandal2017a}, 
\begin{equation}\label{key}
\delta \rho_\mu(\bm{q},t) := \int_{-L/2}^{L/2} \!\text{d}z \!\int_{A}^{} \!\text{d}\bm{r} \exp (\textrm{i}Q_\mu z) e^{\textrm{i}\bm{q}\cdot \bm{r}} \delta  \rho(\bm{r},z,t),
\end{equation}
with the density fluctuations $ \delta\rho(\bm{r},z,t) = \rho(\bm{r},z,t) - n(z) $, in-plane wavevector $\bm{q} =(q_x, q_y)$, discrete wavenumbers $ Q_\mu = 2 
\pi \mu/L, $ and \emph{mode indices} $ \mu \in \mathbb{Z} $. The GSF is then defined as,
\begin{equation}\label{key}
S_{\mu\nu}(q) := \frac{1}{N} \left \langle  \delta \rho_\mu(\bm{q},0)^*  \delta \rho_\nu(\bm{q},0) \right \rangle,
\end{equation}
where $ q = |\bm{q}| $ defines the wavenumber in lateral direction, parallel to the walls. \gj{Here, the average is calculated from many independent initial configurations for which we set $t=0$.}

The matrix-valued GSF $[\mathbf{S}]_{\mu \nu} := S_{\mu\nu}(q)$ can be evaluated theoretically for monodisperse samples using the inhomogeneous density profile $ n(z) $ as input for the generalized Ornstein-Zernicke (OZ) equation \cite{Hansen:Theory_of_Simple_Liquids,Henderson1992,Lang2010,Lang2012},
\begin{equation}\label{key}
\mathbf{S}^{-1} = \frac{n_0}{L^2} \left[ \mathbf{v} - \mathbf{c}  \right].
\end{equation}
Here, $ [\mathbf{c}]_{\mu \nu} = c_{\mu \nu}(q) $ is the direct correlation function, and the inverse-density Fourier amplitudes $ [\mathbf{v}]_{\mu \nu} = v_{\mu - \nu} $ are defined as,
\begin{equation}\label{key}
v_\mu := \int_{-L/2}^{L/2} n(z)^{-1} \exp \left[ {\rm i} Q_\mu z \right] 
\diff z.
\end{equation}
The OZ equation can be solved numerically using a self-consistent iterative procedure based on the Percus-Yerwick closure (OZ+PY). \gj{This procedure takes the density profiles $n(z)$ determined from FMT, as discussed above, as input.} For details on the 
algorithm see Refs.~\cite{Lang2010D,Petersen_2019}. 

To connect our simulation and numerical results to the experiments of Nygård \emph{et al.} \cite{Nygard2009,Nygard2012,Nygard2013,Nygard2016} 
we also calculate an anisotropic structure factor, $ S(q,q_\perp) $, which can be directly measured in scattering experiments, defined as,
\begin{eqnarray}\label{eq:SF1}
S(q, q_\perp) := \frac{1}{N} \left\langle  \delta \hat{\rho}(\bm{q},q_\perp,0)^*  \delta \hat{\rho}(\bm{q},q_\perp,0)  \right\rangle.
\end{eqnarray}
Here, $ q_\perp $ denotes the wavenumber in transverse direction and the fluctuating density modes are given by,
\begin{equation}\label{key}
\delta \hat{\rho}(\bm{q},q_\perp,t) := \int_{-L/2}^{L/2} \text{d}z \int_{A}^{} \text{d}\bm{r} \,  e^{\textrm{i}q_\perp z} e^{\textrm{i}\bm{q}\cdot \bm{r}} 
\delta \rho(\bm{r},z,t).
\end{equation}

\begin{figure}
	\includegraphics[scale=1]{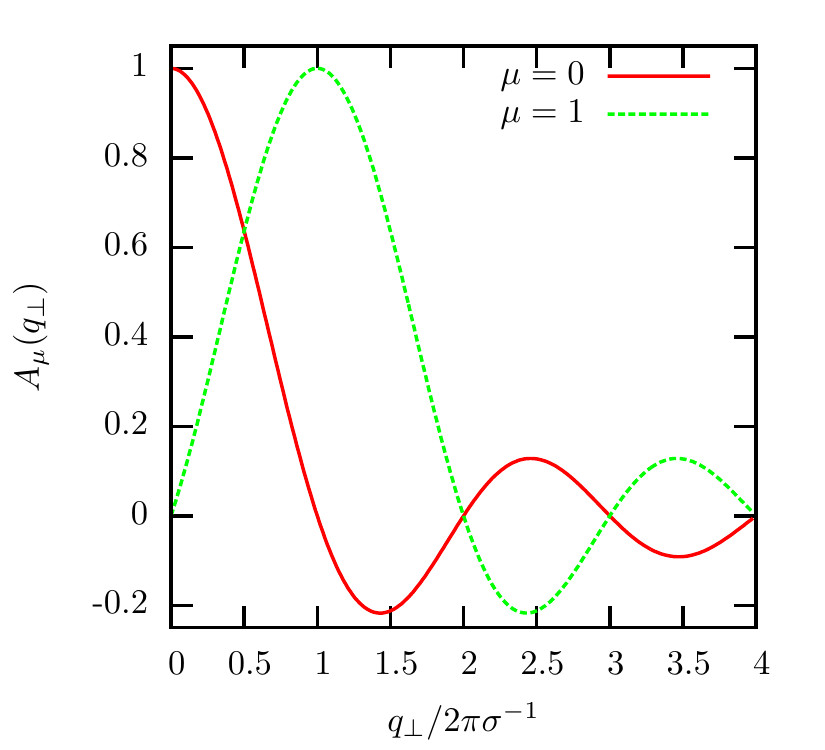}
	\caption{Transverse form factors $ A_\mu(q_\perp) $ defining the transformation from the anisotropic to the generalized structure factor as described by Eq.~(\ref{eq:transformation}). }
	\label{fig:form_factors}
\end{figure}

It should be noted that the definition of the GSF $ S_{\mu \nu}(q) $ contains more information than the experimental definition of $ S(q,q_\perp) $. This is because the off-diagonal components of $ S_{\mu \nu}(q) $, which are non-zero due to the breaking of translational invariance, cannot be calculated from $ S(q,q_\perp) $, since the latter does only contain information on relative 
distances $ z-z^\prime $. Calculating $ S(q,q_\perp) $ from $ S_{\mu \nu}(q) $ is, however, always possible. This can be easily shown by writing the van Hove function, 
\begin{equation}
G(|\bm{r}-\bm{r}^\prime|,z,z^\prime) := n_0^{-1} \left \langle \delta \rho(\bm{r},z,0)  \delta\rho(\bm{r}^\prime,z^\prime,0) \right \rangle,
\end{equation}
in terms of the GSF $S_{\mu \nu}(q) $,
\begin{align}
&G(r,z,z^\prime) =\\ &\int \frac{\text{d} \bm{q}}{(2\pi)^2} \frac{1}{L^2} \sum_{\mu \nu} S_{\mu \nu}(q) \exp \left[ \textrm{i} (Q_\mu z - Q_\nu z^\prime)\right] e^{\textrm{i} \bm{q}\cdot \bm{r}} \nonumber.
\end{align}
This finally enables us to calculate the connection with the anisotropic structure factor $ S(q,q_\perp) $, 
\begin{align}
S(q,q_\perp)& =\int\! \text{d}\bm{r}\! \int_{-L/2}^{L/2}   \! \text{d}z\!    \int_{-L/2}^{L/2} \text{d}z^\prime \, G(r,z,z^\prime) e^{\textrm{i}{q}_\perp (z-z^\prime)}e^{\textrm{i}\bm{q} \cdot r} \nonumber\\
& = \sum_{\mu \nu}  S_{\mu \nu}(q) A_\mu(q_\perp) A_\nu(q_\perp)^*, \label{eq:transformation}
\end{align}
using the transverse form factors,
\begin{equation}\label{eq:form_factors}
A_{\mu}(q_\perp) = \frac{1}{L}  \int_{-L/2}^{L/2}  \text{d}z    e^{\textrm{i}z({q}_\perp-Q_\mu)} = \text{sinc} \left( \frac{L}{2}(q_\perp - Q_\mu) \right),
\end{equation}
with the "sinc" function $\text{sinc}(x) := \sin(x)/x$.
In particular, we find that $ S_{00}(q) = S(q,q_\perp=0)  $ and $ S_{11}(q) = S(q,q_\perp={2 \pi}/{L})  $. More generally, the dependence of the anisotropic structure factor $S(q,q_\perp)$ on values of $ q_\perp 
$ which are not integer multiples of $2\pi/L$, is given by 
a superposition of the individual modes of $ S_{\mu \nu}(q) $ with an amplitude defined by the transverse form factors, $ A_\mu(q_\perp) $ (as visualized in Fig.~\ref{fig:form_factors}).

Although the description of the structure via the GSF $S_{\mu \nu}(q)$ is 
more general than via $S(q,q_\perp)$, the superposition discussed above indicates that some of the information visible in $S(q,q_\perp)$, such as the subtleties of the confinement-induced orientational alignment described in Ref.~\cite{Nygard2009}, could be hidden in the GSF. It therefore appears to be promising to study both definitions of the structure factor. 

\section{Considerations on the convergence towards the 2D fluid}
\label{sec:theory}

A particularly interesting question when analyzing structural properties of strongly confined fluids is their convergence towards their 2D counterparts. In Ref.~\cite{Lang2014a} this problem has been discussed theoretically \gj{for monodisperse systems} and predictions for the rapidity of convergence have been derived which we recapitulate below.

For the specific case of symmetric \gj{and neutral} hard walls the authors derive the asymptotic behavior of the inhomogeneous density profile for constant area number density $ n_0 $ \cite{Lang2014a},
\begin{equation}\label{eq:n_quasi2D}
n(z;L) = \frac{n_0}{L} \left[ 1 + \pi (n_0 L^2) C \left[ \left(\frac{z}{L}\right)^2 - \frac{1}{12} \right] + \mathcal{O}(n_0L^2)^2 \right],
\end{equation}
where, $ C = g(\sigma_+) $ corresponds to the contact value of a 2D hard-disk system with number density $ n_0 $.  It is therefore expected that the density profile becomes parabolic at very small distances and the curvature of the parabola goes to zero in the limit $ L\rightarrow 0, $ if $n(z) L/n_0$ is plotted versus $z/L$.
 This 
parabolic shape has already been confirmed via computer simulations \cite{antonchenko1984nature,doi:10.1063/1.457334}. An interesting feature of the above equation is that 
it allows determining the contact value of a 2D hard-disk fluid, purely by investigating density profiles of a confined hard sphere fluid, as has been pointed out in Ref.~\cite{Lang2014a}. 

The most important results in Ref.~\cite{Lang2014a} concern the rapidity of convergence of the generalized structure factor $ S_{\mu \nu}(q) $ for very small wall 
separations,
\begin{equation}\label{eq:asymptotic_sf}
S_{\mu \nu}(q;L) = \left \{  \begin{array}{ll}
S_\text{2D}(q) \left[ 1 + \mathcal{O}(L^2) \right] & \text{for } \mu=\nu=0\\
\delta_{\mu \nu} + \mathcal{O}(L^2) & \text{else,}\\
\end{array} \right.
\end{equation}
for constant area density $ n_0 $. Here, $ S_\text{2D}(q) $ denotes the structure factor of hard disks at area density $ n_0. $ The equations predict a quadratic convergence of the in-plane GSF $ S_{00}(q) $ towards $ S_\text{2D}(q) $ and similarly state that the generalized structure factor becomes diagonal, but only the in-plane-mode has a non-trivial $ q $-dependence to lowest order in $ L. $

\section{Inhomogeneous Density Profiles}
\label{sec:density}

\begin{figure}
	\includegraphics[scale=0.98]{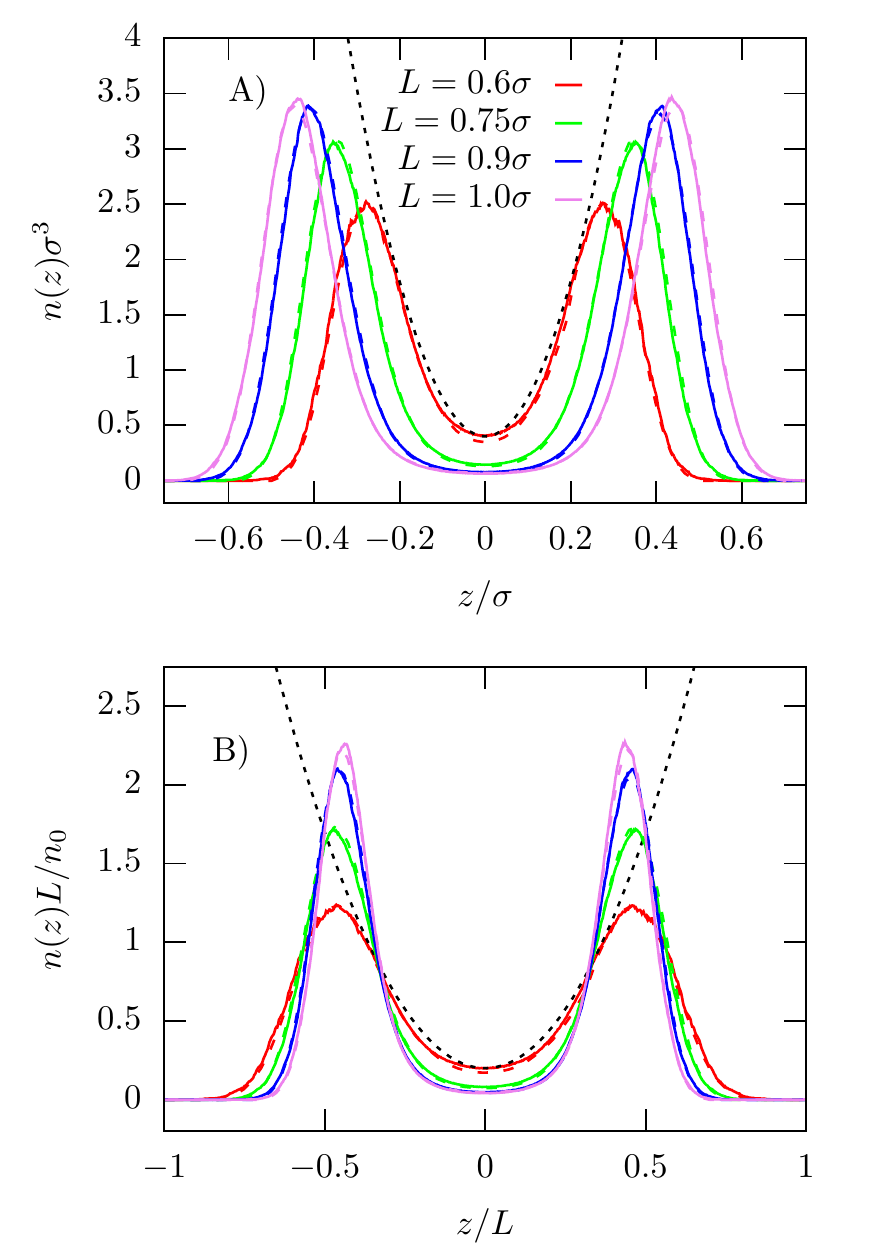}
	\caption{Inhomogeneous density profile $ n(z) $ for polydisperse ($ \delta = 0.15 $) hard spheres at constant packing fraction $ \varphi=0.4 $ for accessible slit widths $ L/\sigma=1.0, 0.9, 0.75, $ and 0.6. Shown are simulations (solid lines) and FMT results (dashed lines) which strongly overlap. Profiles are shifted along the $ y $-axis for the sake of visibility. The black dotted line represents a parabola. A) and B) show the same data in different representation. }
	\label{fig:density_poly}
\end{figure}

\begin{figure}
	\includegraphics[scale=1]{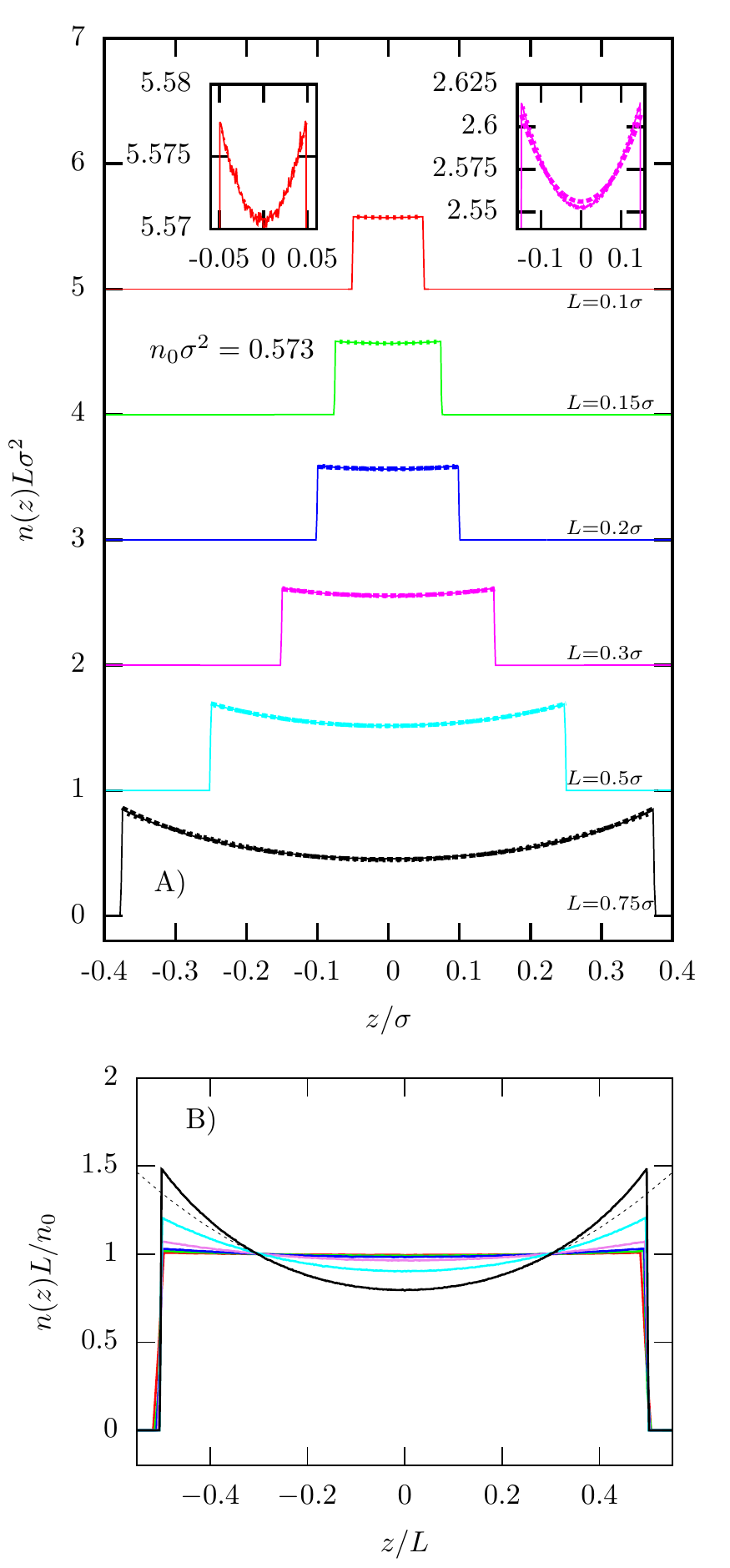}
	\caption{Inhomogeneous density profile $ n(z) $ for monodisperse hard spheres at constant area density $ n_0=0.573 \sigma^{-2} $ for accessible 
slit widths $ L/\sigma=0.75, 0.5, 0.3, 0.2, 0.15 $ and 0.1 (bottom to top). A) Shown are simulations (solid lines), FMT (dashed lines) and quasi-2D 
theory (\ref{eq:n_quasi2D}) (dotted lines) which strongly overlap. Profiles are shifted along the $ y $-axis for the sake of visibility. The insets show a zoom for $ L=0.1\sigma $ (left) and $ L=0.3\sigma $ (right). 
For $ L\leq0.15\sigma $ no convergence was achieved for the FMT calculations. B) Similar simulation data but rescaled according to Eq.~(\ref{eq:n_quasi2D}). The dashed line shows a parabolic law fitted to $L=0.75\sigma$ between $-0.3 \leq z/L \leq 0.3.$ }
	\label{fig:density_n0}
\end{figure}

We first analyze the inhomogeneous density profiles of polydisperse ($ \delta = 0.15 $) hard-sphere liquids for accessible slit widths $ L \leq 1.0\sigma $. At $ L = 1.0\sigma  $ the polydisperse profile reveals a structure with two very pronounced layers and a minimum in the center in which the density is vanishingly low (see Fig.~\ref{fig:density_poly}). When decreasing the wall separation at constant packing fraction $\varphi$, the two layers become slightly less pronounced and their distance is reduced. Importantly, the density in the center minimum increases and one can observe the emergence of a parabolic profile (black, dotted line in Fig.~\ref{fig:density_poly}). The agreement between the FMT results and the 
EDMD simulations is very good. \gj{When plotting the data using dimensionless quantities in Fig.~\ref{fig:density_poly}B) it can be observed that the density profiles indeed start to approach the asymptotic behavior suggested by Eq.~(\ref{eq:n_quasi2D}). Due to the polydispersity it is, however not possible to systematically approach the limit $L \rightarrow 0$ since we are restricted to $L/\sigma \gtrsim
	 0.6$, which is also the minimal system size studied in the experiments \cite{Nygard2016}. } To analyze even smaller wall separations, 
we \gj{therefore} focus on monodisperse liquids in the following. 

\gj{In monodisperse samples, at an accessible slit width $ L = 0.75\sigma $, the density profiles of the monodisperse samples are already approximately parabolic and the agreement of the profiles obtained from simulations, FMT and the asymptotic prediction in Eq.~(\ref{eq:n_quasi2D}) is very good, as shown in Fig.~\ref{fig:density_n0}.  There are only small deviations from the parabolic profile at the boundary (see Fig.~\ref{fig:density_n0}B).} By fixing the area density $ n_0 = 0.573 \sigma^{-2} $ and reducing the wall separation we observe a monotone flattening of the parabolic profile which converges towards a homogeneous density inside the slit in the quasi-two-dimensional limit. These observations are in agreement with results presented in Ref.~\cite{antonchenko1984nature}. Very interestingly, when fixing the packing fraction $ \varphi $ instead of the area density $ n_0 $ one can observe that the agreement between simulations 
and the asymptotic prediction is significantly worse (see Appendix \ref{app:const_phi_density} and Fig.~\ref{fig:density}). This is mainly caused by the denser packing, since $ n_0 = 0.573 \sigma^{-2} $ corresponds to a packing fraction of just $ \varphi = 0.2 $ at $ L=0.5\sigma $, \gj{compared to $\varphi = 0.3$ in Fig.~\ref{fig:density}}. This behavior can already be anticipated from the asymptotic calculation in Eq.~(\ref{eq:n_quasi2D}) which predicts \gj{that the relevant ``smallness'' parameter is not $L/\sigma$ but rather $n_0L^2$. For the profile shown in Fig.~\ref{fig:density} with $L=0.5\sigma$ the smallness is therefore $n_0L^2 = 0.215$ which is thus effectively ``larger'' than for the profiles in Fig.~\ref{fig:density_n0} with $L=0.5\sigma$ which corresponds to $n_0L^2=0.143.$} This analysis therefore shows again that the effects of confinement are strongly enhanced for denser packings \cite{Jung:2020_self}.

\begin{figure}
	\includegraphics[scale=0.9]{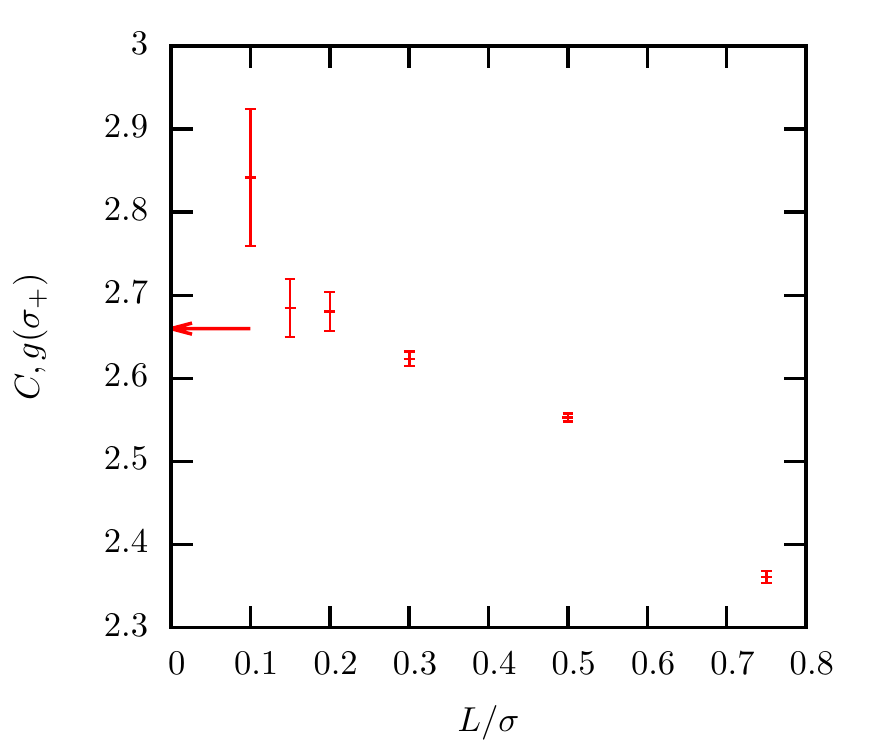}
	\caption{Prefactor $ C $ as defined in Eq.~(\ref{eq:n_quasi2D}) determined by fitting the density profiles for $n_0 \sigma^2 = 0.573$ shown in Fig.~\ref{fig:density_n0}. The arrow marks the contact value of hard disks as determined from simulation results of the radial distribution function $ g(\sigma_+) $.  }
	\label{fig:contact}
\end{figure}

For the above determination of the asymptotic prediction, the contact value $ g(\sigma^+) $ of the 2D hard-disk liquid has been extracted directly 
from EDMD simulations. As discussed in the introduction, one could also invert this calculation and determine the contact value by fitting the profiles shown in Fig.~\ref{fig:density_n0} and extracting the unknown factor $ C $, as defined in Eq.~(\ref{eq:n_quasi2D}). Following this route indeed allows us to determine the contact value with a precision of roughly 2-3\% (see Fig.~\ref{fig:contact}).

\section{Structure Factors and Compressibility}
\label{sec:structure}

\begin{figure}
	\includegraphics[scale=1.0]{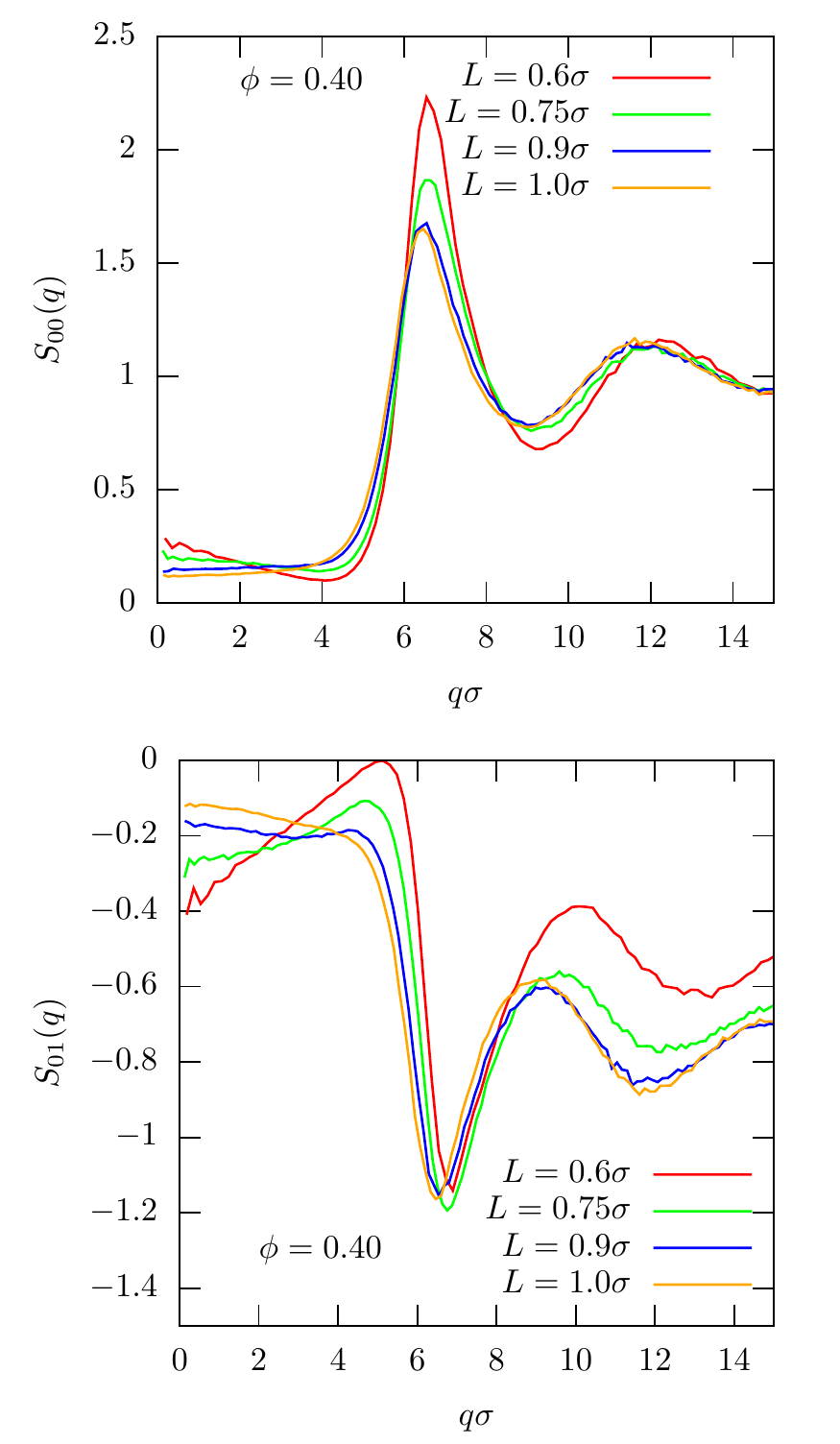}
	\caption{GSF $S_{00}(q)$ (top panel) and $S_{01}(q)$ (bottom panel) for polydisperse ($ \delta = 0.15 $) hard spheres at constant packing fraction $ \varphi=0.4 $. }
	\label{fig:gsf_poly}
\end{figure}

The packing of the particles in confinement can be best described using the generalized and anisotropic structure factors introduced in Section~\ref{sec:structure_intro}. The anisotropic structure factor of polydisperse 
samples in the range $ 0.6 \lesssim L/\sigma \lesssim 1.0 $ has been extensively discussed in Refs.~\cite{Nygard2009,Nygard2016}, here, we will therefore concentrate on the generalized structure factor.

\begin{figure*}
	\includegraphics[scale=1.5]{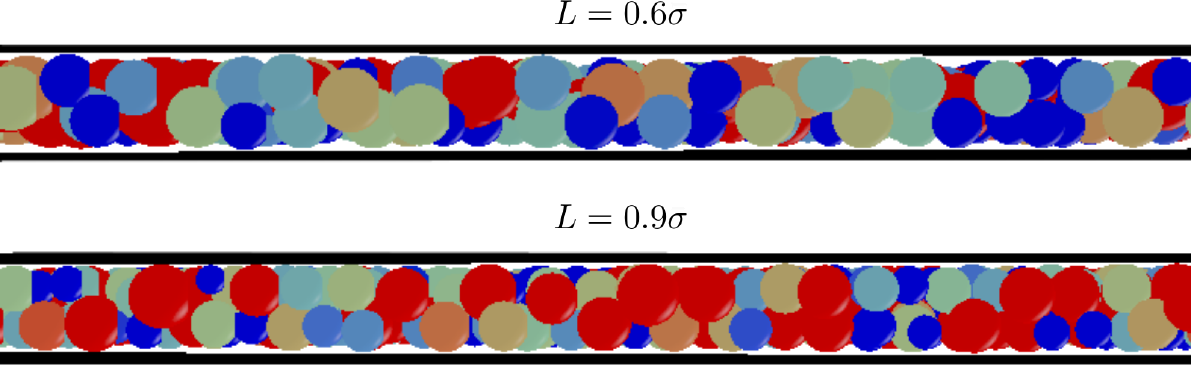}
	\caption{Snapshot of the slit geometry for two accessible slit widths $ L. $ The color code indicates the diameter of the particles (from diameter $  < 0.9\,\sigma $ (blue) to diameter $ > 1.1\,\sigma $ (red) and interpolation between). The snapshot has 
been created with the visualization tool of DynamO \cite{edmd:bannerman2011}. \gj{For $L=0.6\sigma$ many pairs of small spheres with diameter smaller than $\sigma$ can be found (blue color), while at $L=0.9\sigma$ several pairs of small and large spheres with diameter larger than $\sigma$ emerge (red and blue color). }}
	\label{fig:snapshot}
\end{figure*}

When decreasing the wall separation from $ L=1.0\sigma $ to $ L=0.6\sigma $ we observe that the peaks of the in-plane GSF $ S_{00}(q) $ become more pronounced (see Fig.~\ref{fig:gsf_poly}). This finding 
is consistent with the results discussed in Ref.~\cite{Mandal2014} showing that at incommensurate packing (half-integer multiples of $ \sigma $) the peak height is larger than at commensurate packing (integer multiples of $ \sigma $). This non-monotonic dependence can be understood by the interlock of the layered structure for incommensurate packings due to the large amount of particles that have to be squeezed between those layers \cite{Mandal2017a}.

\gj{The off-diagonal generalized structure factor $S_{01}(q)$ is negative indicating that the particles are more likely to be found at the boundary than at the center of the cell. This is a direct consequence of the layering as discussed in Sec.~\ref{sec:density}. Apart from the different sign, the behavior of the off-diagonal is very similar to $S_{00}(q).$ In particular, the minimum at $q\sigma \approx 2 \pi$ for $L=0.6\sigma$ seems to be more pronounced, considering that for this confinement length the off-diagonal component $S_{01}(q)$ converges to a smaller absolute value in the limit $q\rightarrow \infty.$  }

The most striking observation visible in all modes of the generalized structure factor is the \gj{extremum} at small wavenumbers $ q \sigma \approx 4.5 $ for incommensurate packing \gj{(most pronounced for the red curve at $L=0.6\sigma$}). Usually structure factors show a monotonic decay for $q\to 0$ towards a finite value which can be related to the isothermal compressibility  $\kappa_T $ \cite{Hansen:Theory_of_Simple_Liquids}. Similar shapes of the structure factor with minima have been observed in jammed polydisperse samples \cite{Berthier2011}, sticky hard-spheres \cite{Salgi1993}  or 
at macroscopic demixing transitions \cite{Hansen:Theory_of_Simple_Liquids,Binder2010,Hofling2016}, albeit in the latter case the behavior is only visible in the concentration-fluctuation structure factor. We believe that in the present case of confinement the unusual behavior of the structure factor is caused by \emph{micro-demixing}. In Ref.~\cite{Jung2020_cryst} we have studied in detail the confinement-induced demixing and crystallization of polydisperse hard spheres. In the present case, the packing fraction is much smaller, which prevents crystallization but specific wall separations could still favor specific particle sizes. For example, accessible widths of $ L=0.9\sigma $ enable the transverse packing of one small sphere and one large sphere (or two medium sized-spheres), while at $ 
L=0.6\sigma $ only two small spheres can be combined. When studying snapshots of particles packed in slit geometry we indeed observe such a micro-demixing (see Fig.~\ref{fig:snapshot}). We quantify this phenomenon in Appendix~\ref{app:micro_demixing} by calculating the species-dependent radial distribution function and find that incommensurate packing indeed favors the pairing of small spheres and therefore induces a microscopic separation of particles (see Fig.~\ref{fig:gr_demixing} in the Appendix).  The microscopic demixing also occurs at $ L=1.6\sigma $ and for smaller polydispersities $ \delta=0.1 $ and $ \delta = 0.05 
$, but, as could be expected, it is much less pronounced than at extreme confinement and strong polydispersity, as shown in Fig.~\ref{fig:gsf_demixing} in the Appendix. 

\begin{figure}
	\includegraphics[scale=1]{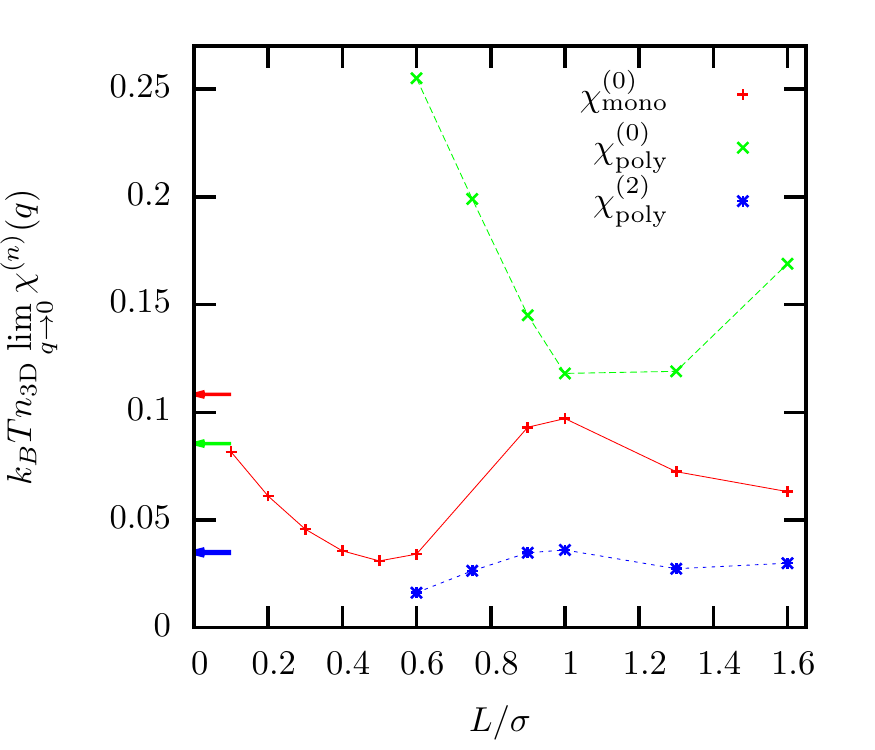}
	\caption{Compressibility as calculated from the small-wavenumber limit of the in-plane GSF Eq.~(\ref{eq:compress}). 
Data is shown for polydisperse samples ($\delta=0.15$) \gj{at} $ \varphi=0.4 $, as characterized in Fig.~\ref{fig:gsf_poly}, and for monodisperse 
samples at $ \varphi=0.3 $, as shown in Fig.~\ref{fig:gsf_mono_constphi}. The arrows show the compressibility of the corresponding two-dimensional fluids at density $ n_\text{2D} = 6 \varphi / \pi \sigma^2$.
}
	\label{fig:compressibility}
\end{figure}

From the \gj{in-plane} generalized structure factor \gj{$S_{00}(q)$} we also extract the long wavelength limit \cite{Kirkwood1951,Salgi1993,Hansen:Theory_of_Simple_Liquids,Berthier2011,Nygard2016},
\begin{equation}\label{eq:compress}
 \lim\limits_{q\rightarrow0}  S_{00}(q) = k_B T n_\text{3D} \chi^{(0)}
\end{equation}
by fitting the GSF to a quadratic function, $ S_{00}(q) = S_{00}(q\rightarrow 0) + Aq^2  + \mathcal{O}(q^4)$ for small wavenumbers. While for monodisperse bulk samples it is well known that $\chi^{(0)}$ corresponds to the isothermal compressibility $\chi_T$ \cite{Hansen:Theory_of_Simple_Liquids} for polydisperse samples it is important to 
realize that $\chi_T$ can only be determined by explicitly taking into account the partial structure factors of the different components \cite{Kirkwood1951,Salgi1993,Berthier2011}. Here we employ a perturbative expression suggested in Ref.~\cite{Berthier2011} to calculate a higher-order quantity $ k_B T n_\text{3D}\chi^{(2)}(q) $ which allows extracting a lateral compressibility $ \chi_T \approx \lim\limits_{q\rightarrow0} \chi^{(2)}(q) $ \gj{from the in-plane GSF}. Details are presented in Appendix \ref{app:pert}.

\begin{figure}
	\includegraphics[scale=1]{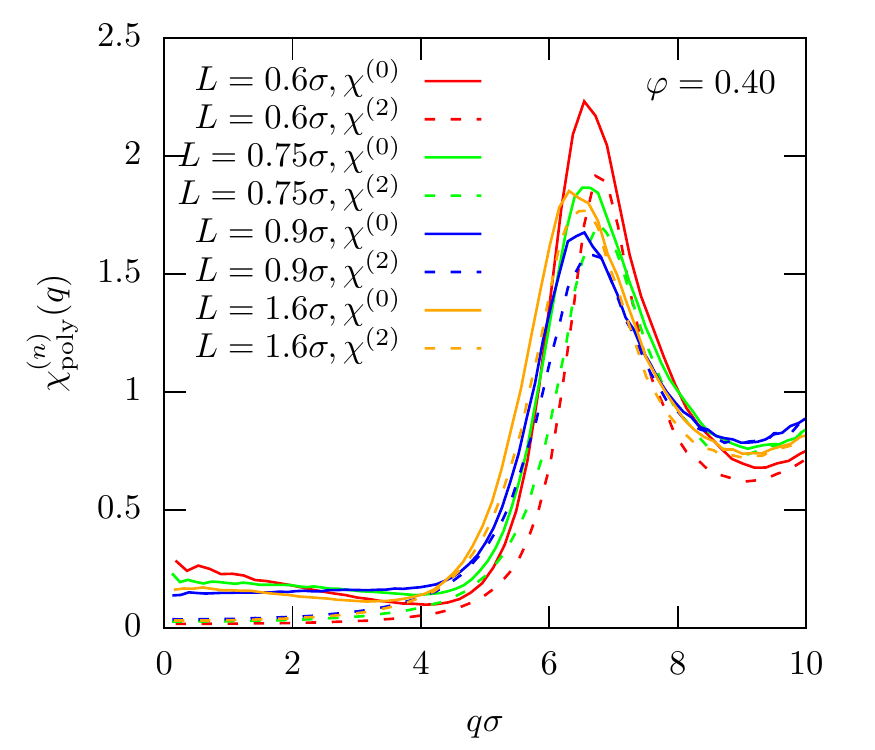}
	\caption{In-plane GSF $ S_{00}(q) = k_B T n_\text{3D} \chi^{(0)}_\text{poly}(q) $ and higher order correction $k_B T n_\text{3D} \chi^{(2)}_\text{poly}(q)$ to account for the polydispersity of the system \cite{Berthier2011}. Data is shown for polydisperse samples ($\delta=0.15$) at $ \varphi=0.4 $.} 
	\label{fig:chi}
\end{figure}

 The results feature a clear non-monotonic dependence of the compressibility (see Fig.~\ref{fig:compressibility}).  Very interestingly one can observe 
that the difference $ \chi_\text{poly}^{(0)}-\chi_\text{poly}^{(2)} $ is much larger for incommensurate packing and actually leads to a trend reversal:  While the naive expression $ \chi_\text{poly}^{(0)} $ would imply that the polydisperse sample has a large compressibility peaking at incommensurate packing, the corrected formula $ \chi_\text{poly}^{(2)} $ reveals that the compressibility in the poly- and monodisprse case behave qualitatively similar and are mainly different due to the different packing fraction. This observation thus highlights the subtle effects of polydispersity and microscopic ordering on the fundamental properties of confined 
liquids. Our simulation result is in agreement with experimental results for confined polydisperse fluids ($\delta=0.12$) \cite{Nygard2016} which also feature an increased long-wavelength limit of the anisotropic structure factor 
$ S(q \rightarrow 0 ) $ at $ L=0.6\sigma $ compared to $ L=1.0\sigma $ (see Fig.~8 in Ref.~\cite{Nygard2016}). In Ref.~\cite{Nygard2016} the effect is explained by the slightly ``frustrated'' structure at incommensurate packing which supposedly leads to an increased compressibility. Our simulations do not support this explanation, since monodisperse spheres feature an inverted behavior although they are clearly also 
affected by the same frustration at incommensurate packing (see Fig.~\ref{fig:compressibility}). From our monodisperse simulations and the correction scheme we therefore conclude that the ``frustrated'' structure effectively reduces the compressibility since the individual layers are interlocked by the particles which are located between those layers.  This effect is further highlighted in Fig.~\ref{fig:chi} in which the original in-plane GSF is compared to the higher-order correction $\chi_\text{poly}^{(2)}(q)$ \cite{Berthier2011} accounting for the polydispersity of the system. The figure shows that the first peak in the structure factor at $ q\sigma \approx 6.5 $ is basically unaffected by the correction, and still exhibits the same non-monotonic behavior as the original structure factor. However, it is clearly visible that the anomalous behavior at small wavenumbers vanishes in the higher-order correction. 

\begin{figure}
\includegraphics[scale=1]{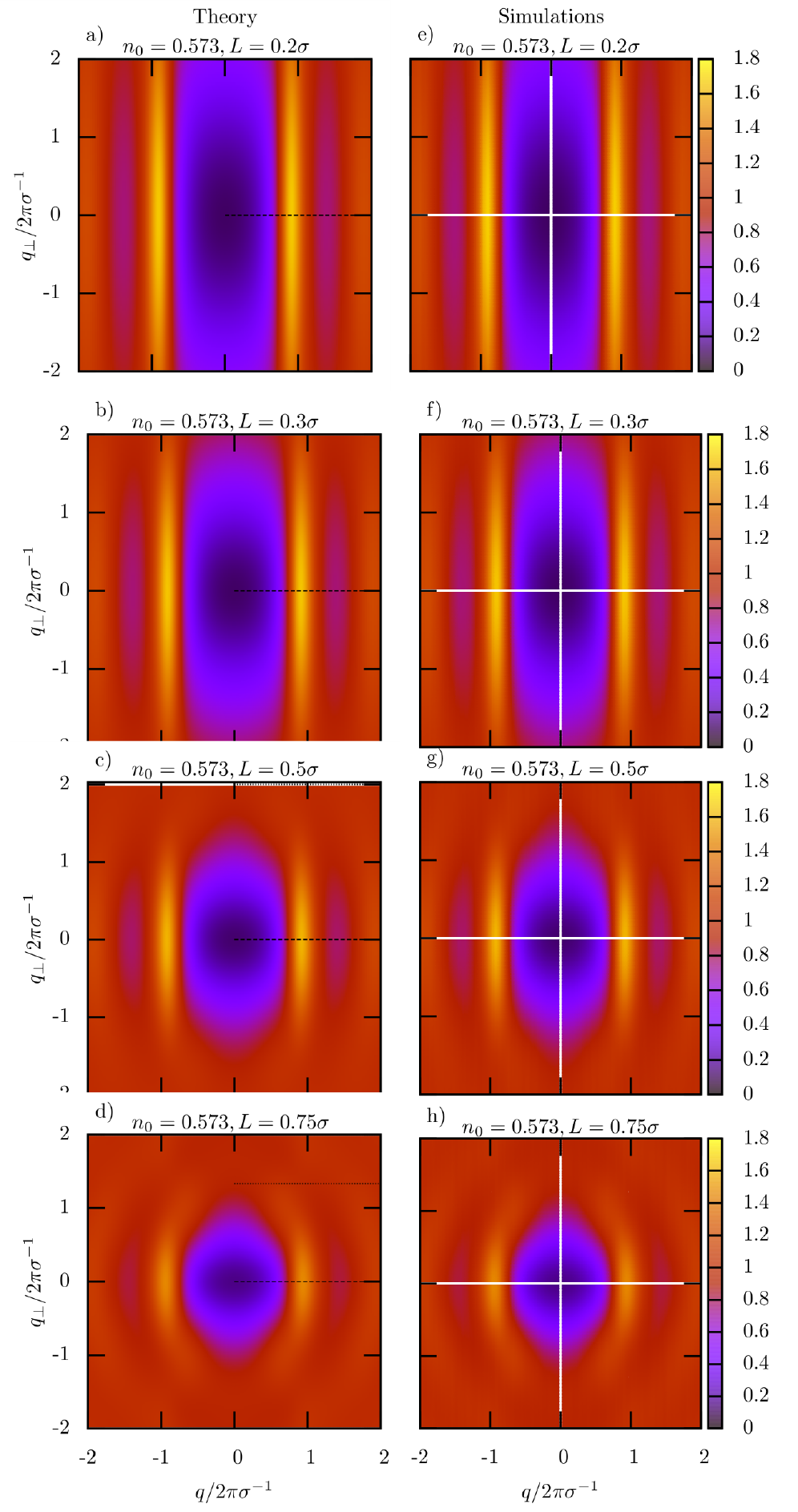}
	\caption{Color plots of the anisotropic structure factor $ S(q,q_\perp) $ for monodisperse hard spheres as calculated from Eq.~(\ref{eq:SF1}), visualized as a function of the wave-vector components in lateral ($ q $) and transverse ($ q_\perp $) direction to the confining walls.  Shown are theory results (FMT\gj{+OZ+PY}) and simulations (EDMD) for wall separations $ L/\sigma = 0.2, 0.3, 0.5 $ and $0.75$. The black dashed and dotted lines highlight the positions of $ S_{00}(q) $ and $ S_{11}(q) $, respectively.}
	\label{fig:SF_anisotropic}
\end{figure}

In the following, we will continue to analyze monodisperse fluids, which allows us to systematically study the limit $L \rightarrow 0$. We first investigate the anisotropic structure factor as defined in Eq.~(\ref{eq:SF1}). For an accessible wall separation $ L=0.75\sigma $ we observe the effect of confinement-induced orientational alignment \cite{Nygard2009,Nygard2016}, visible by the non-spherical profiles displaying increased 
in-plane packing ($ q \sigma = 2 \pi  $, $ q_\perp \sigma = 0  $) and 
basically no  packing in transverse direction ($ q \sigma = 0  $, $ q_\perp \sigma = 2 \pi  $) (see Fig.~\ref{fig:SF_anisotropic}). In fact, Fig.~\ref{fig:SF_anisotropic}d is very similar to Fig.~5b in Ref.~\cite{Nygard2016}. Upon further reduction of the wall separation, the most prominent observation is that the originally circular center becomes increasingly stretched until the dependence on $ q_\perp $ becomes very weak for $ L=0.2\sigma $, indicating the approach of the quasi-two-dimensional limit.  This transformation is very well described by both OZ+PY and computer simulations.

\begin{figure}
	\includegraphics{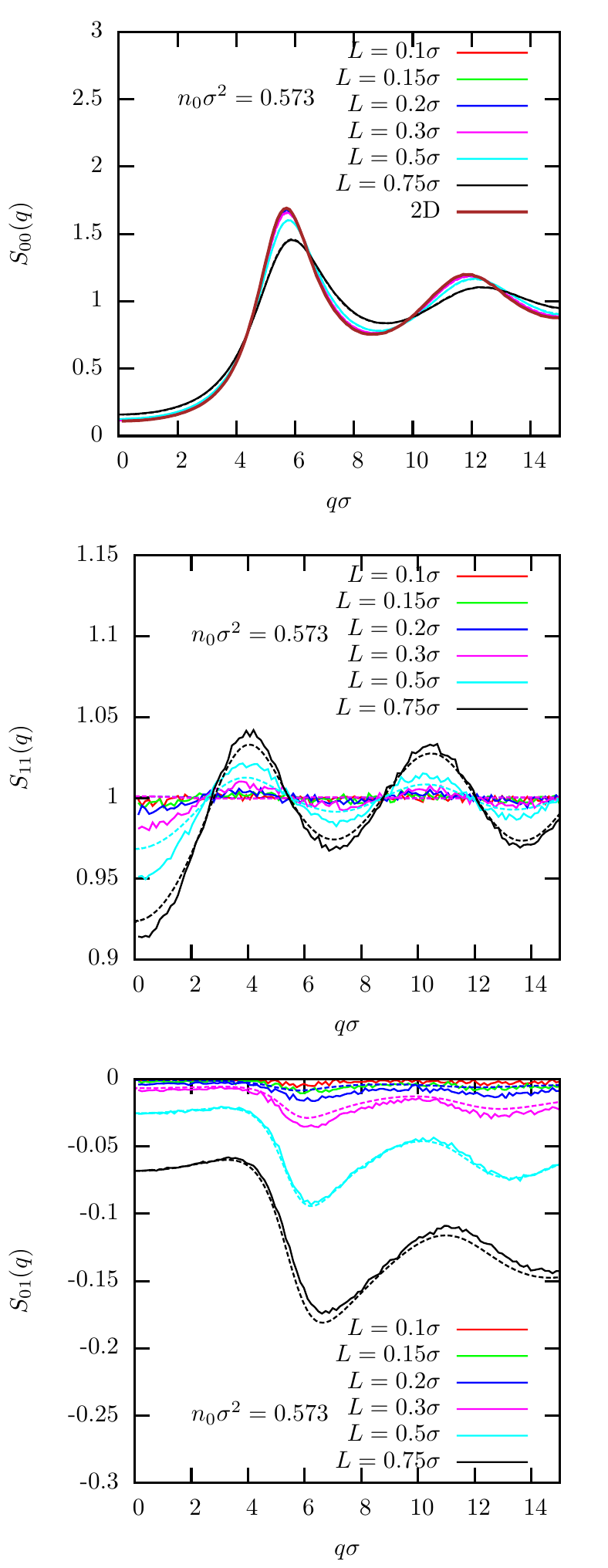}
	\caption{GSF $ S_{\mu \nu}(q) $ for monodisperse hard spheres at constant area density $ n_0=0.573 $ for modes $ \nu=\nu=0 $ (top panel), $ \mu=\nu=1 $ (middle panel) and $ \mu=0, \nu=1 $ (bottom panel). Results are shown for computer simulations (full line) and FMT\gj{+OZ+PY} (dashed line). For $S_{00}(q)$ the results of theory and simulations overlap. }
	\label{fig:SF_generalized}
\end{figure}

To study the convergence towards the 2D limit more quantitatively we calculate the generalized structure factor $ S_{\mu \nu}(q) $. In agreement with the asymptotic relations given in Eq.~(\ref{eq:asymptotic_sf}) we indeed observe that the in-plane structure factor converges smoothly towards 
its 2D counterpart, $ S_{00}(q) \to S_\text{2D}(q) $, as shown in Fig.~\ref{fig:SF_generalized}, which also corresponds to the profile $ S(q,q_\perp =0) $ as visible in Fig.~\ref{fig:SF_anisotropic}. Similarly, the off-diagonal components are vanishing in the 2D limit, $ S_{\mu \nu}(q) \to \gj{0} $ for $ \mu \neq \nu $, and the remaining diagonal terms converge towards \gj{$ S_{\mu \mu}(q)\to 1 $ (for $\mu > 0$).} The latter can already be anticipated from Fig.~\ref{fig:SF_anisotropic}c,d since it corresponds to the profile $ S(q,q_\perp= 2 \pi \mu /L) $  and thus already lies outside the circular center for $ L=0.5\sigma, $ despite the observed stretching of the profiles in the $ q_\perp $-direction. When comparing theory and simulations one can observe a very good agreement. The only exception is the GSF $ S_{11}(q) $, where one can observe that FMT\gj{+OZ+PY} underestimates the oscillations around $ S_{11}(q) = 1 $. These deviations are, however, very small in absolute terms.

\begin{figure}
	\includegraphics[scale=1]{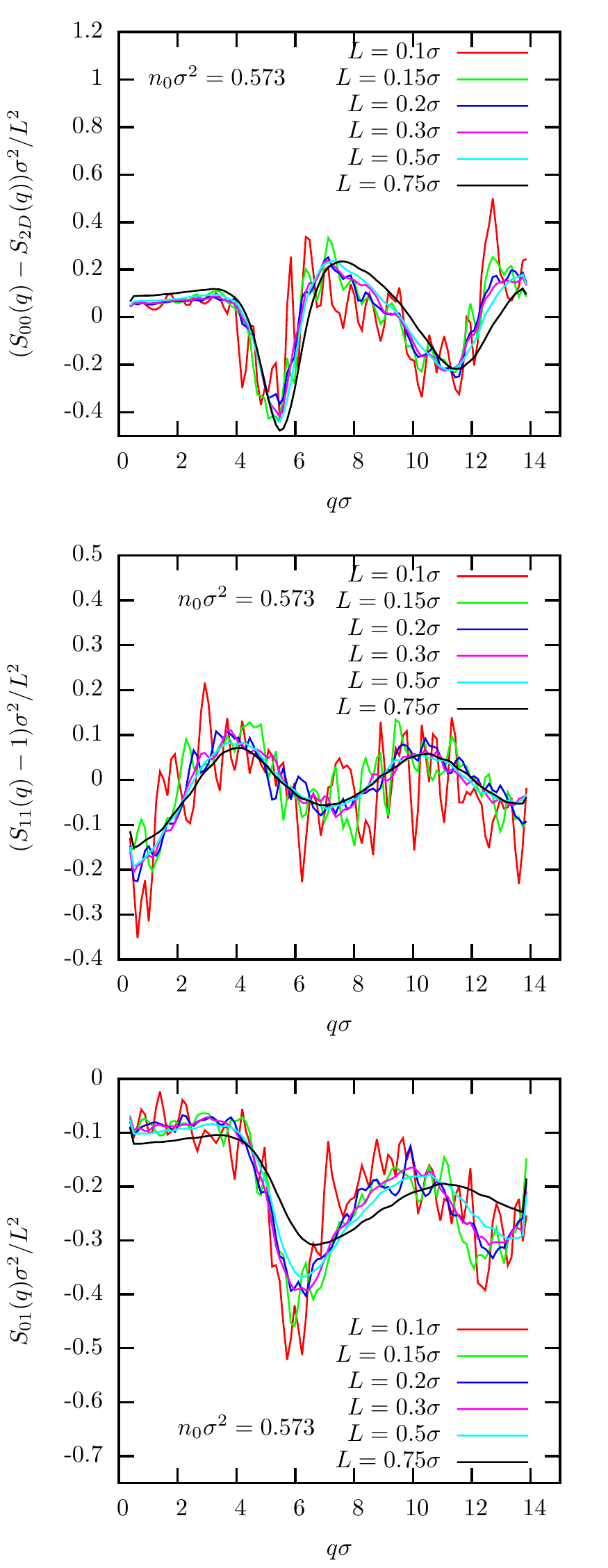}
	\caption{Difference between the generalized structure factor $ S_{\mu \nu}(q) $ for monodisperse hard spheres at constant area density $ n_0=0.573 $ and the 2D limit. The difference are scaled by $ L^{-2} $.}
	\label{fig:sf_convergence}
\end{figure}

We can now analyze the convergence towards the 2D limit by subtracting the respective limits and dividing by the theoretically predicted quadratic 
power of convergence. By doing this, we find the important result that the quadratic convergence can indeed be confirmed by the EDMD simulations, since all curves collapse onto a single master curve  (see Fig.~\ref{fig:sf_convergence}). Interestingly, these master curves are comparable in shape 
for the in-plane mode $ S_{00}(q) $ and the first off-diagonal mode $ S_{01}(q) $, while  $ S_{11}(q) $ shows completely different convergence behavior. From Fig.~\ref{fig:sf_convergence} we can also assess the impact of 
higher-order terms on the GSF. While the diagonal modes seem to be well described by the asymptotic formulas up to an accessible slit width of $ L=0.75\sigma $ the off-diagonal mode already shows significant deviations for this wall separation. 

\begin{figure}
	\includegraphics[scale=1]{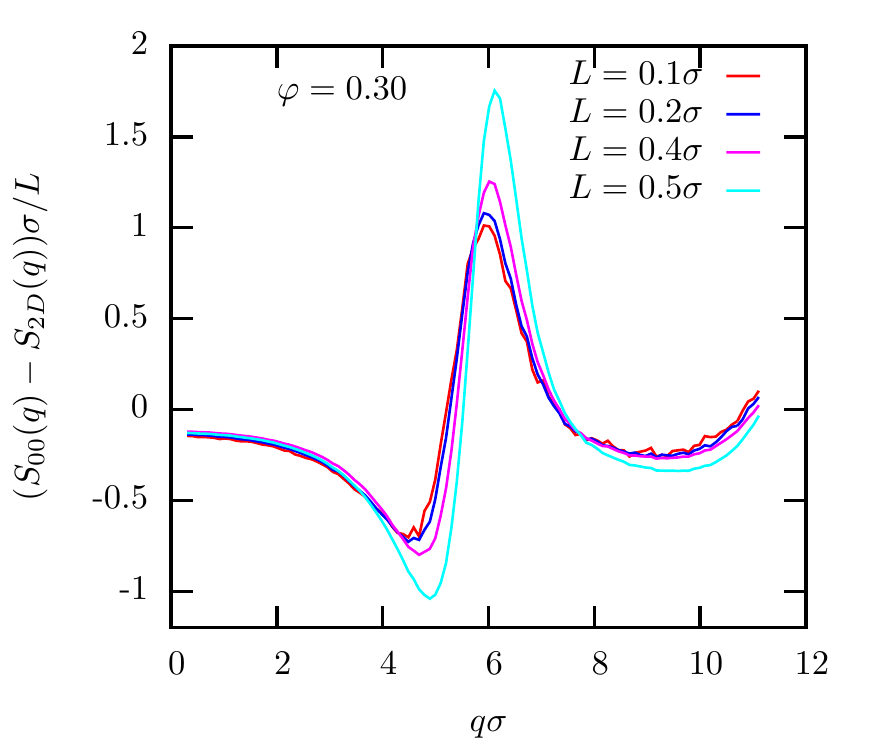}
	\caption{Differences between the generalized structure factor $ S_{00}(q) $ for monodisperse hard spheres at constant packing fraction $ \varphi=0.3 $ and the corresponding 2D limit. The differences are scaled by $ L^{-1} $.}
	\label{fig:SF_00}
\end{figure}

 It should be noted that the quadratic convergence strongly depends on which control parameter is kept constant. The asymptotic theory predicts a quadratic convergence for constant $ n_0 $, which could be confirmed by computer simulations. However, when keeping the packing fraction $ \varphi 
$ constant, the in-plane mode surprisingly shows a linear convergence, as 
can be clearly observed in Fig.~\ref{fig:SF_00} (observe the different rescaling factor). This change of the power of convergence clearly originates from the linear scaling with $ L $ in the relationship between the area density $ n_0 $ and the packing fraction $ \varphi $,
 \begin{equation}\label{key}
n_0\frac{\pi \sigma^2}{6}  =   \left(1 + \frac{L}{\sigma} \right) \varphi.
\end{equation}
Consequently, by choosing the packing fraction as control parameter, one loses one order in the convergence rate although the asymptotic limit 
is the same. The other modes are not affected by the control parameter and also show quadratic convergence for constant $ \varphi $ as is shown in Appendix \ref{app:const_phi} and Fig.~\ref{fig:SF_converge_phi}.

\section{Summary and Conclusions}
\label{sec:conclusions}

We have investigated structural properties of poly- and monodisperse hard 
spheres in extreme confinement by fundamental{-}measure theory and computer simulations. For polydisperse samples we found an interesting microscopic demixing effect, which can be seen as precursor of the confinement-induced crystallization through demixing reported in Ref.~\cite{Jung2020_cryst} and has direct consequences on the qualitative behavior of the structure factor. We have further shown that the application of a perturbation expansion is essential to calculate the compressibility from the structure factor in order 
to account for the impact of polydispersity. 

While polydisperse hard spheres prevent the approach of the two-dimensional limit we found for the inhomogeneous density profiles of monodisperse hard-sphere liquids the expected emergence of parabolic profiles which smoothly converge towards flat profiles in the two-dimensional limit. Importantly, we observe very good agreement  between the  asymptotic theory \cite{Lang2014a} and simulations which opens up the possibility to extract the contact value of hard-disk liquids from the data. This is a fascinating result since the contact value is, 
for example, connected to the pressure of the hard-disk system via the equation of state \cite{Hansen:Theory_of_Simple_Liquids}, showing that one can determine thermodynamic quantities of a two-dimensional system by observing density profiles in confined geometry.   We also studied extensively the convergence of the 
anisotropic and generalized structure factors towards their two-dimensional counterparts. As predicted by the asymptotic theory, we found a smooth 
quadratic convergence if the area density is kept constant, as opposed to 
a linear convergence of the in-plane mode for constant packing fraction. 

Our work significantly enhances the understanding of simple liquids in confinement by smoothly connecting their structural properties in extreme confinement to the known structural properties of two-dimensional liquids. 
This result is also particularly important with respect to the understanding of dynamical properties in this regime since we know that the impact 
of confinement on diffusion and glass-transition is strongly connected to 
the structural quantities \cite{PhysRevLett.96.177804,Mittal:PRL100:2008,Mandal2014,Nygard2017,Jung:2020}. In future work it would therefore be interesting to extend the present analysis to dynamical properties such as diffusion coefficients and decorrelation time scales \cite{Mittal:PRL100:2008,Mandal2017,PhysRevE.95.032604,Schilling2016}, the critical packing fraction for the glass transition \cite{Mandal2014} or the non-ergodicity parameters \cite{Mandal2017a} and investigate whether they similarly smoothly approach the properties of their two-dimensional counterparts.

 It would also be exiting to validate whether a scenario in which the structure factor attains a minimum at small $ q $ is also observable in experiments. Very precise experimental data has been presented in Ref.~\cite{Nygard2016}  for a particle dispersity of roughly $ \delta=0.12 $. While these curves neither confirm nor disprove the existence of such a minimum, one can clearly see that at incommensurate packing in their Fig.~8a, the curves are leveling off much stronger than in their Fig.~8b for commensurate packing and it is definitely possible that they grow in the regime $ q \sigma 
\lesssim  2 $ where no experimental data is available. It should also 
be discussed whether the impact of the higher-order expansion to calculate the compressibility from the structure factor in polydisperse samples \cite{Berthier2011}, as extensively discussed in this manuscript, might also affect the interpretation of the experiments in Ref.~\cite{Nygard2016}. Since a straightforward application of the algorithm in Ref.~\cite{Berthier2011} requires knowledge of every particle diameter it will, however, be challenging to apply the same correction to experimental measurements. 
 
 \section*{Acknowledgments} 
 
We thank Rolf Schilling for constructive criticism on the manuscript. This work has been supported by the Austrian Science Fund (FWF): I 5257-N. 
 
  \FloatBarrier
 
 \appendix
 
  \section{Asymptotic convergence at constant packing fraction}
 \label{app:const_phi}
 
 We have discussed in the main text that the convergence towards the two-dimensional hard-disk fluid can be best described using a constant area density $ n_0 $ as control parameter instead of the packing fraction $ \varphi. $ Here, we will present some results for constant packing fraction to supplement and substantiate this discussion.
 
 \subsection{Inhomogeneous density profile}
 \label{app:const_phi_density}
 
  \begin{figure}
 	\includegraphics[scale=1]{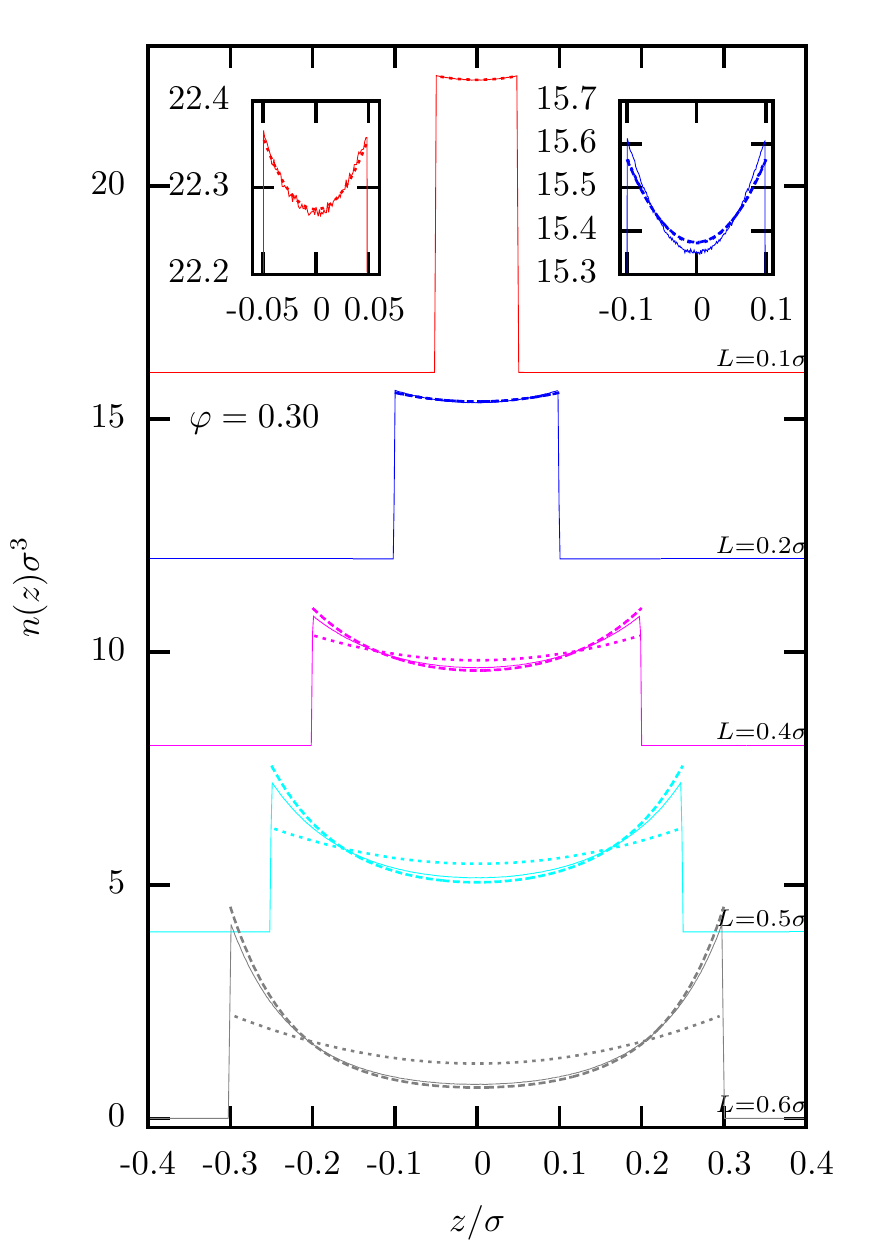}
 	\caption{Inhomogeneous density profile $ n(z) $ for monodisperse hard spheres at constant packing fraction $ \varphi=0.3 $ for accessible slit 
widths $ L/\sigma=0.6, 0.5, 0.4, 0.2 $ and 0.1 (bottom to top). Shown are simulations (solid lines), FMT (dashed lines) and quasi-2D theory (\ref{eq:n_quasi2D}) (dotted lines). Profiles are shifted along the $ y $-axis for the sake of visibility. The insets show a zoom of the $ L/\sigma=0.1 $ (left) and $ L/\sigma=0.2 $ (right). For $ L/\sigma=0.1 $ the FMT result did not converge, while for $ L/\sigma=0.2 $ the FMT and quasi-2D results perfectly overlap (even in the inset).}
 	\label{fig:density}
 \end{figure}

 When studying the inhomogeneous density profile at constant packing fraction for various wall separations, one observes that the agreement between EDMD simulations and the asymptotic theory \cite{Lang2014a} is much worse than what we have seen for constant area density $ n_0 $ (compare Figs.~\ref{fig:density_n0} and \ref{fig:density}). The first reason is that 
the profiles at constant $ \varphi  $ at larger accessible slit widths $ L $ are measured for much denser systems than the ones shown in Fig.~\ref{fig:density_n0} for constant area density. One consequence of this is that at $ L=0.6\sigma 
$ the profile is not even perfectly parabolic. The second reason is that the density of the two-dimensional reference fluid which we use to determine $ C=g(\sigma^+) $ is not well defined. Here, we used the one with $n_{\text{2D}} = 6 \varphi/\pi \sigma^2$, but one could also argue that $ n_\text{2D} = 6 \varphi (L+\sigma) /\pi\sigma^3 = n_0 $ would be a reasonable choice. This would lead to a larger curvature of the parabolic profiles and thus certainly improve the agreement between computer simulations and asymptotic theory in Fig.~\ref{fig:density}. This shows that a constant $ n_0 $ is indeed the proper control parameter to investigate the approach to the two-dimensional limit.
 
 \subsection{Structure factor}
 
   \begin{figure}
	\includegraphics[scale=1]{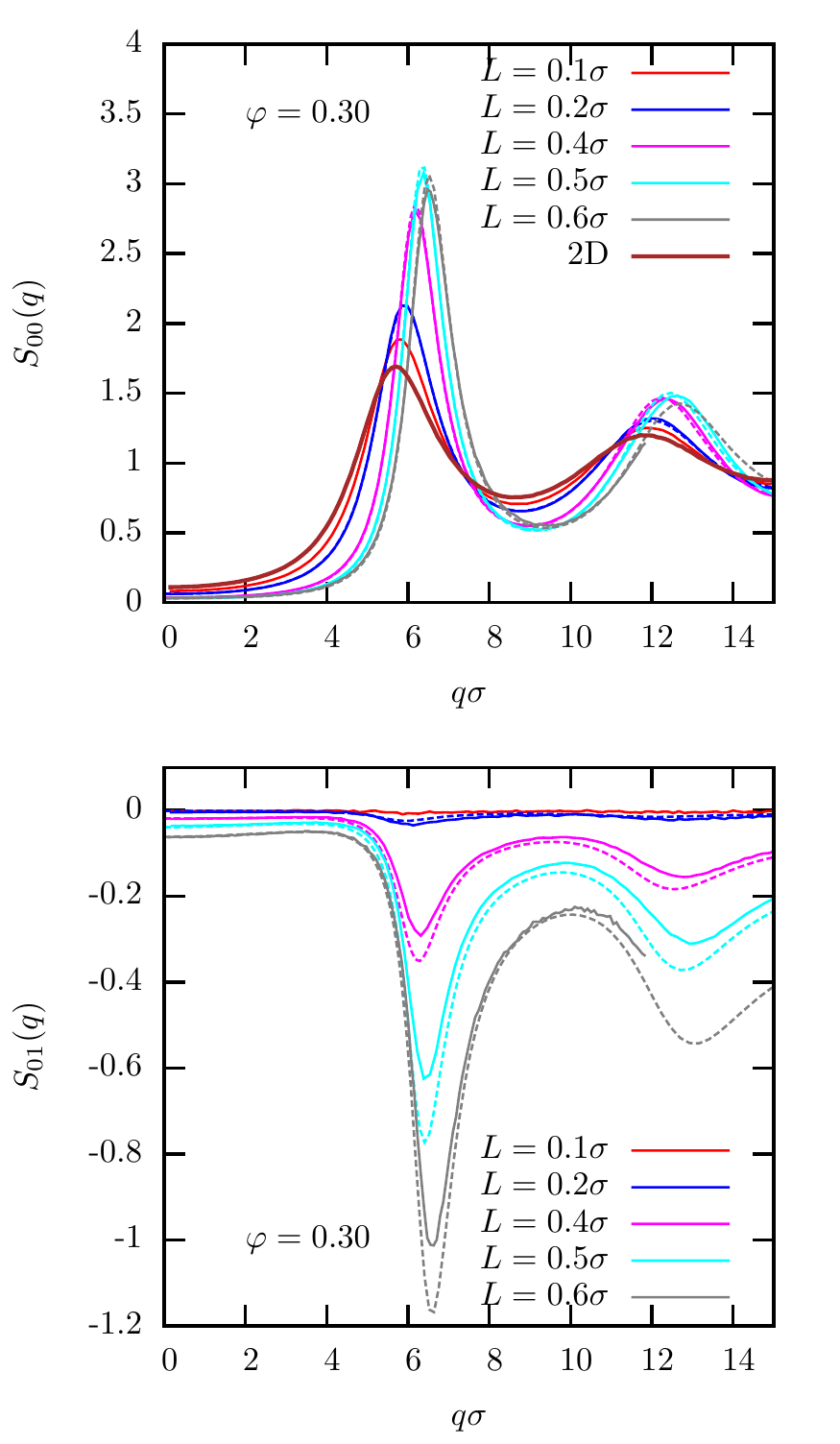}
	\caption{GSF $ S_{\mu \nu}(q) $ for monodisperse hard spheres at $ \varphi=0.3 $. Results are shown for computer simulations (full 
line) and FMT\gj{+OZ+PY} (dashed line).}
	\label{fig:gsf_mono_constphi}
\end{figure}

 \begin{figure}
	\includegraphics[scale=1]{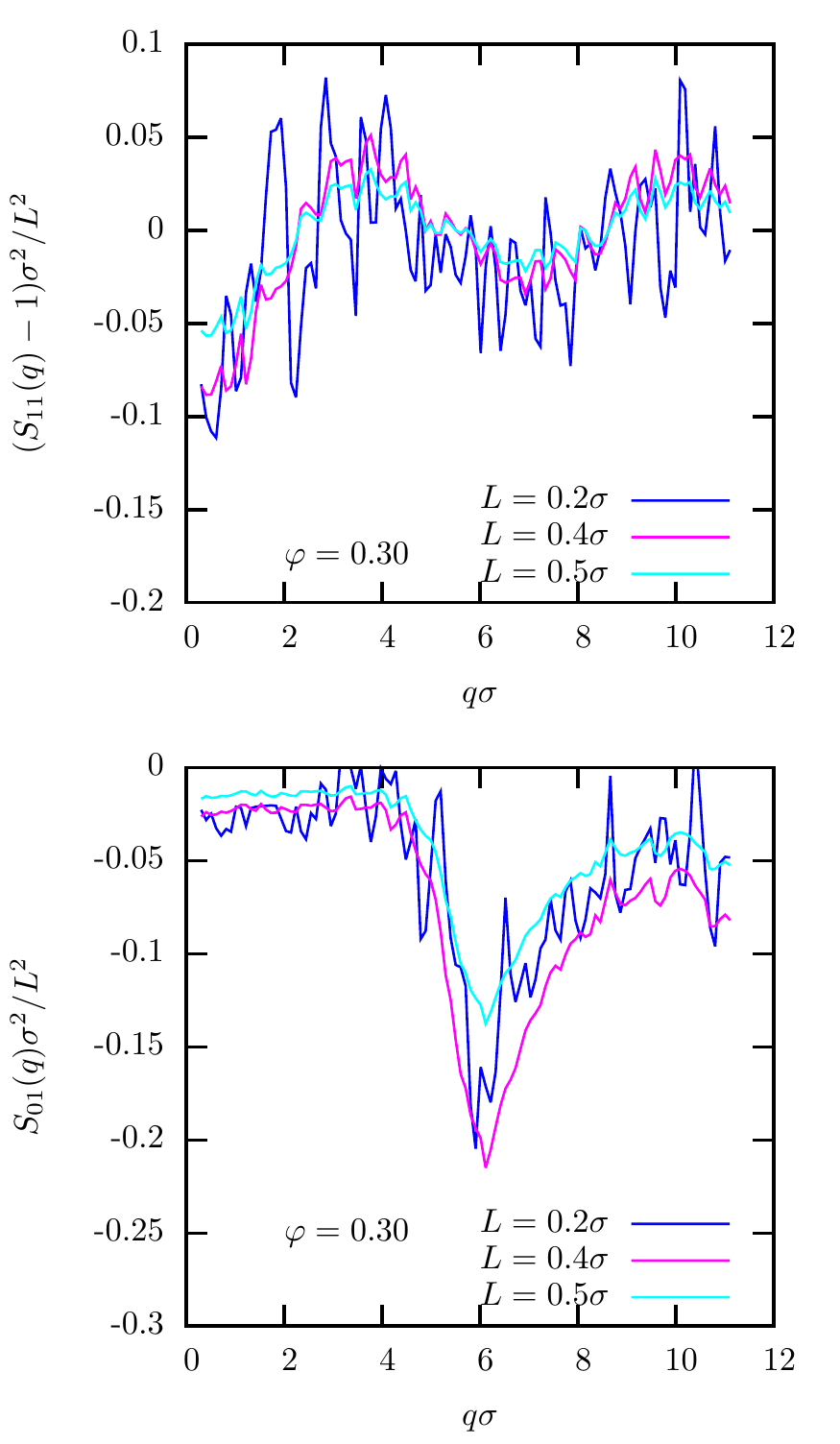}
	\caption{Differences between the GSF $ S_{\mu \nu}(q) $ for monodisperse hard spheres at $ \varphi=0.3 $ and the corresponding 2D limit. The differences are scaled by $ L^{-1} $ ($ \mu=\nu=0 $) and $ L^{-2} $ otherwise.}
	\label{fig:SF_converge_phi}
\end{figure}

The convergence of the generalized structure factor towards its two-dimensional counterpart for constant packing fraction is qualitatively different than what 
we have discussed for constant area density $ n_0 $ (compare Figs.~\ref{fig:SF_generalized} and \ref{fig:gsf_mono_constphi}). The most important difference between constant area density $ n_0 $ and packing fraction $ \varphi $ is the change in the order of convergence in the in-plane structure factor, as has already been discussed in the main text (see Section~\ref{sec:structure} and Fig.~\ref{fig:SF_00}). This observation only holds for the in-plane mode, since the higher-order modes still show a quadratic convergence (see Fig.~\ref{fig:SF_converge_phi}). Also for these modes, 
however, the range in which this asymptotic behavior can be observed is significantly marginalized and valid only up to $ L \lesssim 0.3 \sigma. $

One can also observe that the peak height of the in-plane structure factor decreases upon reduction of the accessible slit width, starting from $ L=0.5\sigma. $ This is consistent with earlier observations in the range $ 1.0 \lesssim L/\sigma\lesssim 2.0 $ in Refs.~\cite{Mandal2014,Mandal2017} where it was pointed out that at incommensurate packing the peak height of the generalized structure factor indeed attains a maximum. The agreement between FMT\gj{+OZ+PY} and computer simulations is still relatively good, but different from the 
data shown in the main text clear deviations can be observed. This observation can be explained with the higher density of the samples, which directly affects the precision of the FMT excess free energy $ \mathcal{F}^\text{ex}[n_i] $, as defined in Eq.~(\ref{eq:fmt}), since it is approximated via a 
low-density expansion  \cite{fmt:Rosenfeld1989,fmt:Roth2010}.

  \section{Microscopic demixing}
 \label{app:micro_demixing}
 
    \begin{figure}[b]
 	\includegraphics[scale=1]{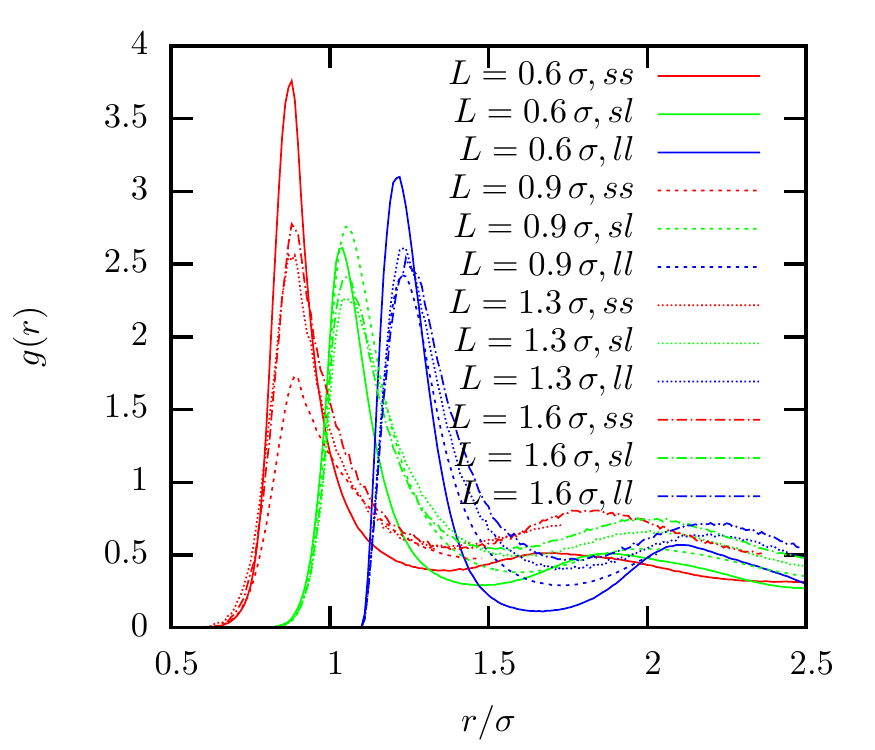}
 	\caption{Radial distribution function $g(r)$ for polydisperse ($ \delta 
= 0.15 $) hard spheres at constant packing fraction $ \varphi=0.4 $, calculated for different pairs of species: small-small (ss), small-large (sl) and large-large (ll). All particles with diameter $ d < 0.9\sigma $ ($ d > 1.1\sigma $) are accounted to the small (large) species, respectively.  }
 	\label{fig:gr_demixing}
 \end{figure} 

  \begin{figure}
	\includegraphics[scale=1]{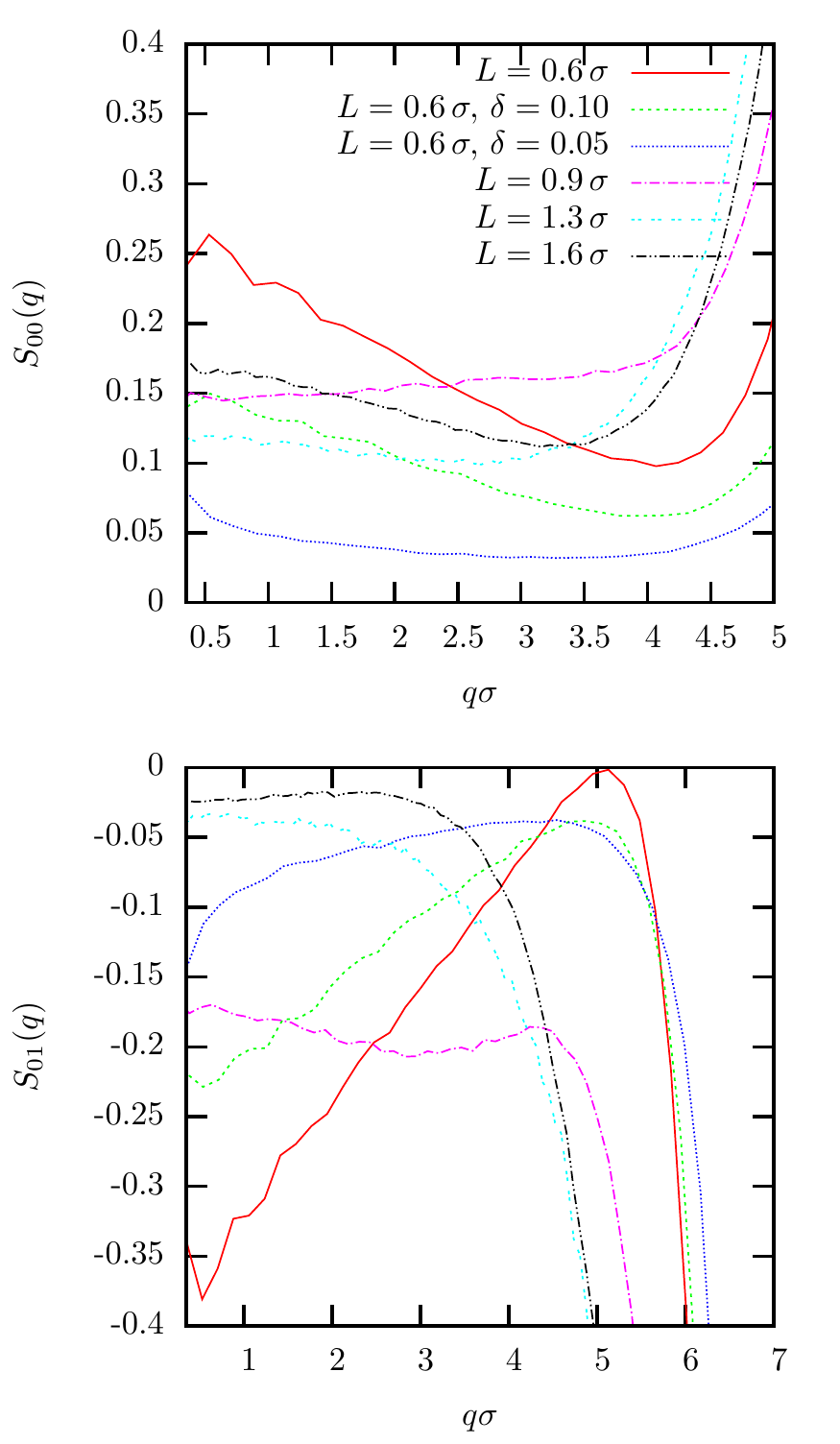}
	\caption{GSF $ S_{\mu \nu}(q) $ for polydisperse ($ \delta = 0.15 $) hard spheres at constant packing fraction $ \varphi=0.4 $ for small wavenumbers $ q\sigma<5 $. For $ L=0.6\sigma $, the figure also shows data for polydispersity $ \delta=0.10 $ (green, dashed) and $ \delta=0.05 $ (blue, dotted), respectively. }
	\label{fig:gsf_demixing}
\end{figure}
 
 To explain the qualitatively different shape of the GSFs for polydisperse hard spheres compared to monodisperse samples at incommensurate packing, as visible in Fig.~\ref{fig:gsf_poly} we have introduced in the main text the notion of microscopic demixing. This effect describes 
the preferred pairing of similarly sized spheres on a microscopic level, and should not be confused with a \emph{macroscopic} demixing transition \cite{Hansen:Theory_of_Simple_Liquids}.

The microscopic demixing can already be anticipated from the snapshot in Fig.~\ref{fig:snapshot} clearly showing at $ L=0.6{\sigma} $ the tendency to combine two small spheres to fill the slit, as opposed to $ L=0.9{\sigma} $ where on average a small sphere is rather combined with a large sphere. We quantify this observation by calculating the radial distribution function,
\begin{equation}\label{key}
g_{a b}(r) = \left\langle\frac{V}{ 4 \pi r^2 N_a N_b} \sum_{i \in N_a, j \in N_b}  \delta(|\bm{R}_i(t)-\bm{R}_j(t)| - r)\right\rangle,
\end{equation}
for particles that are \gj{either smaller ($s$) or larger ($l$) than} $\sigma$, $a,b \in \{ s,l\} $.  Here, $ N_a $ and $ N_b $ denote 
the total number of particles of species $ a $ and $ b $, respectively, and $ \bm{R}_i(t)=(x_i(t),y_i(t),z_i(t))  $ the particle positions. The radial distribution function clearly shows that at incommensurate packing 
($L=0.6\sigma$) the small and large particles tend to be coupled ($ g_{ss}(r) $ and $ g_{ll}(r) $ have the highest peak), while at commensurate packing ($L=0.9\sigma$) the mixing term $ g_{sl}(r) $ features the highest peak (see Fig.~\ref{fig:gr_demixing}). We believe that this demixing then leads to the significant increase of the long-wavelength limit of the structure factor, which finally induces the minimum at $ q\sigma\approx 
4 $, as shown in Fig.~\ref{fig:gsf_demixing}.  We emphasize that this 
effect does not translate to the compressibility which actually shows a minimum for incommensurate packing despite the simultaneous increase of $ S_{00}(q) $ at small wavenumbers as discussed in Section~\ref{sec:structure}.

\section{Perturbative expansion to calculate compressibility in polydisperse fluids}
\label{app:pert}

It is well known that the long-wavelength limit of the structure factor of monodisperse bulk liquids is directly connected to the isothermal compressibility \cite{Hansen:Theory_of_Simple_Liquids},
\begin{equation}
  \lim\limits_{q\rightarrow0}  S(q) = k_B T n_\text{3D} \chi_T.
\end{equation}
This formula, however, does not hold anymore for polydisperse systems. Given a system with $ M $ different species it has been derived that the isothermal compressibility can be determined as \cite{Kirkwood1951,Salgi1993},
 \begin{equation}\label{key}
 \left[n_\text{3D}k_B T \chi_T\right]^{-1} =  \lim\limits_{q\rightarrow0} \bm{x}^T \bm{\tilde{S}}^{-1}(q)\bm{x},
 \end{equation}
 where \gj{$\left[\bm{\tilde{S}}(q)\right]_{ij} = N^{-1} \left\langle \tilde{\rho}_i(\bm{q})^*  \tilde{\rho}_j(\bm{q}) \right\rangle$} denotes partial structure factors for the individual species, $i,j=1,\ldots, M$, \gj{$\tilde{\rho}_i(\bm{q})$ are the partial density modes}, and $ \bm{x}^T=(x_1,\ldots,x_M) $ is the concentration vector with $ x_i = N_i/N. $ \gj{Please note that the partial structure factor $\left[\bm{\tilde{S}}(q)\right]_{ij}$ and the partial densities $\tilde{\rho}_i$ should not be confused with the modes of the generalized structure factor $S_{\mu \nu}(q)$ and the density modes $\rho_\mu(\bm{q})$ in the main text. }

To apply a similar formalism to continuously polydisperse systems Berthier \emph{et al.} \cite{Berthier2011} have derived a perturbative expansion based on the moment-density fields $ \epsilon^k(\bm{q}) = \sum_{n=1}^N \epsilon_n^k e^{\textrm{i} \bm{q} \cdot \bm{R}_n} $, defined in terms of powers of the relative diameter deviations $ \epsilon_n=(\sigma_n-\bar{\sigma})/\bar{\sigma} $, with average particle diameter $ \bar{\sigma} $ and the particle position $ \bm{R}_n=(x_n,y_n,z_n)  $. Using the moment structure factors,
\begin{equation}\label{key}
\left[\hat{\bm{S}}_{(\alpha)}(q)\right]_{kl} = N^{-1} \left\langle \epsilon^k(\bm{q})^*  \epsilon^l(\bm{q}) \right\rangle, \quad k,l=1,\ldots,\alpha.
\end{equation}
 one can evaluate the order $ \alpha $ of the expansion in the form,
\begin{equation}
\left[n_\text{3D}k_B T \chi^{(\alpha)}(q)\right]^{-1} = \bm{m}_\alpha^T 
\hat{\bm{S}}_{(\alpha)}^{-1}(q) \bm{m}_\alpha.
\end{equation}
Here, we have defined the moment vector $ \bm{m}_\alpha^T = (\delta_1,\ldots,\delta_\alpha), $ with $ \delta_k = \sum_{n=1}^N x_n \epsilon^k_n  $. In particular we recover the structure factor $ S(q) = n_\text{3D}k_B T \chi^{(0)}(q)$ from the zero-order term.  The compressibility can then be evaluated as $ \chi_T = \lim\limits_{q\rightarrow0} \chi^{(\alpha)}(q) $ for high orders $ \alpha. $

The above expansion can directly be applied to $ S_{00}(q) $ by defining $ \epsilon^k(\bm{q}) = \sum_{n=1}^N \epsilon_n^k e^{\textrm{i} \bm{q} \cdot \bm{r}_n} $ using the in-plane coordinates. In this manuscript 
we have evaluated this expansion to second order $ \alpha=2 $ which was 
reported in Ref.~\cite{Berthier2011} to be applicable up to dispersities of $ \delta_\text{max}= 0.3 $ which is significantly larger than the value of $ \delta=0.15 $ considered in this work.

\FloatBarrier

\bibliography{library_local}

\end{document}